\documentclass[aps,prb,groupedaddress,twocolumn,showpacs]{revtex4-1}

\usepackage{graphicx}
\usepackage{color}

\begin{document}

\newcommand{\fab}[1]{{\bf \textcolor{red}{#1}}}


\title{Scattering nonlocality in quantum charge transport: \\
Application to semiconductor nanostructures}


\author{Roberto Rosati}
\author{Fausto Rossi}
\email[]{Fausto.Rossi@polito.it}
\homepage[]{staff.polito.it/Fausto.Rossi}
\affiliation{
Department of Applied Science and Technology, Politecnico di Torino \\
C.so Duca degli Abruzzi 24, 10129 Torino, Italy
}


\date{\today}

\begin{abstract}

Our primary goal is to provide a rigorous treatment of scattering nonlocality in semiconductor nanostructures. On the one hand, starting from the conventional density-matrix formulation and employing as ideal instrument for the study of the semiclassical limit the well-known Wigner-function picture, we shall perform a fully quantum-mechanical derivation of the space-dependent Boltzmann equation. 
On the other hand, we shall examine the validity limits of such semiclassical framework, pointing out, in particular, regimes where scattering-nonlocality effects may play a relevant role; to this end we shall supplement our analytical investigation with a number of simulated experiments, discussing and further expanding preliminary studies of scattering-induced quantum diffusion in GaN-based nanomaterials. 
As for the case of carrier-carrier relaxation in photoexcited semiconductors, our analysis will show the failure of simplified dephasing models in describing phonon-induced scattering nonlocality, pointing out that such limitation is particularly severe for the case of quasielastic dissipation processes.

\end{abstract}

 

\maketitle

\section{Introduction}\label{s-I}

Since the seminal paper by Esaki and Tsu,\cite{Esaki70a} artificially tailored as well as self-assembled semiconducting nanostructures\cite{b-Ihn10} form the leading edge of semiconductor science and technology.\cite{Liu12a,Cheng13a,LeBlanc13a} The design of state-of-the-art optoelectronic devices, in fact, heavily exploits the principles of band-gap engineering,\cite{b-Capasso11} achieved by confining charge carriers in spatial regions comparable to their de Broglie wavelengths.\cite{b-Bastard88} This, together with the progressive reduction of the typical time-scales involved, pushes device miniaturization toward limits where the application of the traditional Boltzmann transport theory\cite{b-Jacoboni89} becomes questionable, and a comparison with more rigorous quantum-transport approaches\cite{Frensley90a,Axt98a,Datta00a,Rossi02b,Axt04a,Pecchia04a,Iotti05b} is imperative; the latter can be qualitatively subdivided into two main classes. On the one hand, so-called double-time approaches based on the nonequilibrium Green's function technique have been proposed and widely employed; an introduction to the theory of nonequilibrium Green's functions with applications to many problems in transport and optics of semiconductors can be found in the books by Haug and Jauho,\cite{b-Haug07} Bonitz,\cite{b-Bonitz98} and Datta;\cite{b-Datta05}
by employing ---and further developing and extending--- such nonequilibrium Green's function formalism, a number of groups have recently proposed efficient quantum-transport treatments for the study of various meso- and nanoscale structures as well as of corresponding micro- and optoelectronic devices.\cite{Taylor01a,Faleev05a,Luisier09a,Zhang12a}
On the other hand, so-called single-time approaches based on the density-matrix formalism\cite{b-Haug04,b-Rossi11} have been proposed (see Sec.~\ref{s-DMF}), including phase-space treatments\cite{Frensley90a,b-Buot09} based on the Wigner-function formalism (see Sec.~\ref{s-WF}).

In spite of the intrinsic validity limits of the semiclassical theory just recalled, during the last decades a number of Boltzmann-like Monte Carlo simulation schemes have been extensively employed for the investigation of new-generation semiconductor nanodevices.\cite{Ryzhii98a,Iotti01b,Kohler02a,Callebaut04a,Lu06a,Bellotti08a,Jirauschek10a,Matyas10a,Iotti10a,Vitiello12a,Matyas13a,Iotti13a} 
Such modeling strategies ---based on the neglect of carrier phase coherence--- are however unable to properly describe space-dependent ultrafast phenomena. To this aim, the crucial step is to adopt a quantum-mechanical description of the carrier subsystem; this can be performed at different levels, ranging from phenomenological dissipation and decoherence models\cite{Gmachl01a} to quantum-kinetic treatments.\cite{Axt98a,Rossi02b,Axt04a}
Indeed, in order to overcome the intrinsic limitations of the semiclassical picture in properly describing ultrafast space-dependent phenomena ---e.g., real-space transfer and escape versus capture processes--- Jacoboni and co-workers have proposed a quantum Monte Carlo technique,\cite{Brunetti94a} while Kuhn and co-workers have proposed a quantum-kinetic treatment;\cite{Reiter07a} 
however, due to their high computational cost, these non-Markovian density-matrix approaches are often unsuitable for the design and optimization of new-generation nanodevices.

In order to overcome such limitations, a conceptually simple as well as physically reliable quantum-mechanical generalization of the conventional Boltzmann theory has been recently proposed.\cite{Rosati13b} The latter preserves the power and flexibility of the semiclassical picture in describing a large variety of scattering mechanisms; more specifically, employing a microscopic derivation of generalized scattering rates based on a recent reformulation of the Markov limit,\cite{Taj09b} a density-matrix equation has been derived, able to properly account for space-dependent ultrafast dynamics in semiconductor nanostructures; indeed, the density-matrix approach proposed in Ref.~\onlinecite{Rosati13b} has been recently applied to the analysis of genuine quantum-diffusion phenomena in GaN-based bulk and nanostructured materials,\cite{Rosati14a} allowing for a preliminary analysis of free-carrier versus scattering-induced diffusion. 

Primary goal of this paper is to provide a rigorous treatment of scattering nonlocality. On the one hand, starting from the conventional density-matrix formulation\cite{b-Kubo12a,b-Kubo12b} and employing as ideal instrument for the study of the semiclassical limit the well-known Wigner-function picture,\cite{b-Kubo12a,b-Zachos05} we shall perform a fully quantum-mechanical derivation of the space-dependent Boltzmann equation.
On the other hand, we shall examine the validity limits of such semiclassical approximation scheme, pointing out, in particular, regimes where scattering-nonlocality effects may play a relevant role; to this end we shall supplement our analytical investigation with a number of simulated experiments, discussing and further expanding the preliminary study of scattering-induced quantum diffusion in GaN-based nanomaterials recently presented in Ref.~\onlinecite{Rosati14a}.
As for the case of carrier-carrier relaxation in photoexcited semiconductors,\cite{Rossi02b} our analysis will show the failure of simplified dephasing models in describing phonon-induced scattering nonlocality, pointing out that such limitation is particularly severe for the case of quasielastic dissipation processes.

The Paper is organized as follows: In Sect.~\ref{s-DMF} we shall recall and discuss the basic concepts and instruments commonly employed for the microscopic investigation of high-field transport and/or ultrafast optical excitations in semiconductor materials in terms of the single-particle density-matrix formalism.
In Sect.~\ref{s-WF} we shall introduce the well-known Wigner-function picture; the latter ---often regarded as a classical-like phase-space representation of quantum mechanics---
will allows us to identify the general approximation scheme needed in order to derive the conventional space-dependent Boltzmann equation from the density-matrix formalism. 
Thanks to a few prototypical simulated experiments, in Sect.~\ref{s-sid} we shall be able to identify conditions where scattering-nonlocality effects ---absent within the semiclassical treatment--- may play a crucial role.
Finally, in Sec.~\ref{s-SC} we shall summarize and draw a few conclusions.

\section{Fundamentals of the density-matrix formalism}\label{s-DMF}

In order to investigate in fully quantum-mechanical terms the electro-optical response of semiconductor materials and related devices, it is crucial to study the time evolution of single-particle quantities, such as the total carrier density, mean kinetic energy, charge current, and so on. In general, such quantities
are given by a suitable (quantum-plus-statistical) average of a corresponding
(single-particle) operator $\hat{a}$, usually expressed in terms of the single-particle density-matrix operator~$\hat{\rho}$ as\cite{b-Rossi11}
\begin{equation}\label{av1}
\langle a \rangle = {\rm tr}\left\{\hat{a} \hat{\rho}\right\}\ .
\end{equation}
It follows that within the Schr\"odinger picture the crucial step is to analyze the time evolution of the single-particle density-matrix operator $\hat{\rho}$, whose equation of motion is always of the general form:\cite{b-Rossi11}
\begin{equation}\label{DME}
\frac{d \hat{\rho}}{d t} 
= 
\left.\frac{d \hat{\rho}}{d t}\right|_{\rm sp}
+ 
\left.\frac{d \hat{\rho}}{d t}\right|_{\rm scat}\ .
\end{equation}
Here
\begin{equation}\label{DME-epsilon}
\left.\frac{d \hat{\rho}}{d t}\right|_{\rm sp}
=
\frac{1}{i\hbar}\,\left[\hat{H}_{\rm sp},\, \hat{\rho}\,\right]
\end{equation}
describes the coherent dynamics dictated by the noninteracting-electron Hamiltonian $\hat{H}_{\rm sp}$ (including elastic single-electron scattering processes as well as various lowest-order renormalization contributions) while, by neglecting so-called memory effects (see below), 
\begin{equation}\label{DME-Gamma}
\left.\frac{d \hat{\rho}}{d t}\right|_{\rm scat}
= 
\Gamma\,(\hat{\rho})
\end{equation}
is, in general, a non-linear superoperator describing energy dissipation and decoherence that electrons experience within the host material. 

The above single-particle picture has been applied to a variety of physical problems,\cite{b-Rossi11} ranging from quantum-transport phenomena to ultrafast electro-optical processes; however, it is vital to stress that the  degree of accuracy of such density-matrix formalism is intimately related to the choice of the scattering superoperator $\Gamma$ in (\ref{DME-Gamma}). 

The microscopic derivation of suitable scattering superoperators has been one of the most challenging problems in solid-state physics.
Indeed, For purely atomic and/or photonic systems, dissipation and decoherence
phenomena may be successfully described via adiabatic-decoupling
procedures\cite{b-Scully97} in terms of extremely simplified models via phenomenological parameters; within such  effective treatments, the main goal is to identify a suitable form of the Liouville superoperator, able to ensure the positive-definite character of the corresponding density-matrix
operator.\cite{b-Davies76} This is usually accomplished by identifying proper Lindblad superoperators,\cite{Lindblad76a} expressed in terms of a few crucial system-environment coupling parameters.
In contrast, solid-state materials and devices are often characterized by a complex many-electron quantum evolution, resulting in a non-trivial interplay between coherent dynamics and energy-dissipation and decoherence processes;\cite{b-Haug04,b-Rossi11} it follows that for a quantitative description of such coherence-versus-dissipation coupling the latter needs to be treated via fully microscopic approaches.

Based on the pioneering works by Van Hove,\cite{VanHove57a} Kohn and Luttinger,\cite{Kohn57a} and Zwanzig,\cite{Zwanzig61a} a number of adiabatic- or Markov-approximation schemes have been developed and employed for the study of quantum-transport and coherent-optics phenomena in semiconductor materials and devices; the latter may be divided into two general categories: approaches based on semiclassical (i.e., diagonal) scattering superoperators also referred to as Pauli master equations,\cite{Fischetti99a,Gebauer04a,Knezevic08a} and fully quantum-mechanical (i.e., non-diagonal) dissipation models.\cite{Lindberg88a,Kuhn92a,Hohenester97a,BiSun99a,Flindt04a} 
Moreover, in order to account for non-markovian or memory effects ---relevant in the presence of strong couplings and/or extremely short excitations--- a number of quantum-kinetic approaches have been also considered.\cite{TranThoai93a,Schilp94a}

As far as the Markov treatments are concerned, the latter depend strongly on the particular problem under investigation, and therefore the resulting set of kinetic equations describes a specific subsystem of interest, e.g., a gas of $N$ electrons or excitons, a single carrier, etc.
Moreover, as originally pointed out by Spohn and co-workers,\cite{Spohn80a} kinetic approaches based on the conventional Markov limit may lead to the violation of the positive-definite character of the density-matrix operator, and therefore to unphysical results; in particular, they
clearly pointed out that the choice of the adiabatic decoupling strategy is definitely not unique, and only one among the available possibilities, developed in the pioneering work by Davies\cite{b-Davies76}, could be shown to preserve positivity: it was the case of a ``small'' subsystem of interest interacting with
a thermal environment, and selected through a partial-trace
reduction. Unfortunately, this theory was restricted to finite-dimensional subsystems only (i.e., $N$-level atoms), and to the particular projection scheme of the partial trace.

To overcome this serious limitation in the study of solid-state systems, an alternative and more general Markov procedure has recently been proposed;\cite{Taj09b} the latter (i) in the discrete-spectrum case coincides with the Davies model just recalled, (ii) in the semiclassical limit (see below) reduces to the well-known Fermi's golden rule, and (iii) describes a genuine Lindblad evolution also in the continuous-spectrum case, thus providing a reliable and robust treatment of energy-dissipation and decoherence processes in semiconductor quantum devices.
As discussed in Ref.~\onlinecite{Taj09b}, by means of such alternative adiabatic-decoupling approach, different Markovian approximations are generated by choosing different projection schemes (corresponding to different subsystems of interest, e.g., many-electron description, single-particle picture, etc.). However, we stress that, opposite to standard master-equation formulations,\cite{b-Davies76,Spohn80a} in this new adiabatic-decoupling strategy positivity is intrinsic, and does not depend on the chosen subsystem of interest.

As discussed in App.~\ref{App-LSS}, by applying such general adiabatic-decoupling scheme together with the usual mean-field approximation,\cite{Rossi02b} it is possible to perform a microscopic derivation of the single-particle scattering superoperator $\Gamma$ in (\ref{DME-Gamma}); in particular, for any single-particle interaction mechanism it is possible to derive a non-linear scattering superoperator of the form
\begin{equation}\label{NSS}
\Gamma \,(\hat{\rho})
=
\sum_s 
\frac{1}{2} \left(
(\hat\mathcal{I} - \hat\rho) \hat A^s \hat\rho \hat A^{s \dagger}
- 
\hat A^{s \dagger} (\hat\mathcal{I} - \hat\rho) \hat A^s \hat\rho\right)
\, + \textrm{H.c.}\ ,
\end{equation}
where $\hat\mathcal{I}$ is the identity operator and ``H.c.'' denotes the Hermitian conjugate.
As we can see, the non-linear character of the above scattering superoperator originates from the so-called Pauli factors $(\hat\mathcal{I} - \hat\rho)$; indeed, by neglecting such nonlinearities, i.e., $\hat\mathcal{I} - \hat\rho \to \hat\mathcal{I}$, the scattering term in (\ref{NSS}) reduces to the following Lindblad superoperator:
\begin{equation}\label{Lindblad}
\Gamma \,(\hat{\rho})
=
\sum_s \left(
\hat A^s \hat\rho \hat A^{s \dagger}
- 
\frac{1}{2} \left\{\hat A^{s \dagger} \hat A^s, \hat\rho\right\} 
\right) \ .
\end{equation}
It follows that, by neglecting such Pauli factors, for each single-particle interaction mechanism $s$  one is thus able to perform a fully microscopic derivation of a corresponding Lindblad superoperator, thereby preserving the positive-definite character of the single-particle density matrix $\hat\rho$.
The main features of such microscopic treatment are briefly recalled and discussed in App.~\ref{App-LSS}, where we report the explicit form of the Lindblad operators $\hat A^s$ for the relevant case of carrier-phonon interaction.

We stress that, strictly speaking, these Pauli factors vanish in the low-density limit only; however, in this limit the single-particle density-matrix formalism becomes highly questionable, since in this regime electron-hole Coulomb-correlation dominates. It follows that the use of the Lindblad scattering superoperator in (\ref{Lindblad}) is well justified in semiconductor bulk and nanostructured materials characterized by carrier densities sufficiently high to neglect excitonic effects, and sufficiently low to neglect the above non-linear Pauli contributions; as a matter of fact, such requirements are often fulfilled by new-generation semiconductor quantum devices.

By denoting with $| \alpha \rangle$ the eigenstates of $\hat{H}_{\rm sp}$ (corresponding to the energy spectrum $\epsilon_\alpha$), the density-matrix equation (\ref{DME}) can also be written as
\begin{equation}\label{SBE}
\frac{d \rho_{\alpha_1\alpha_2}}{dt} = \frac{\epsilon_{\alpha_1}-\epsilon_{\alpha_2}}{i\hbar}\,\rho_{\alpha_1\alpha_2} +
\left.\frac{d \rho_{\alpha_1\alpha_2}}{d t}\right|_{\rm scat}\ .
\end{equation}
Such set of coupled equations of motion for the density-matrix elements $\rho_{\alpha_1\alpha_2}$ are usually referred to as the semiconductor Bloch equations.\cite{b-Haug04}
In particular, the diagonal elements ($\rho_{\alpha_1 = \alpha_2}$) describe state populations, while non-diagonal contributions ($\rho_{\alpha_1
\neq \alpha_2}$) ---also referred to as inter-state polarizations--- describe quantum-mechanical phase coherence between the single-particle states $\alpha_1$ and $\alpha_2$.\cite{b-Rossi11}

By adopting as scattering superoperator the Lindblad-like prescription in (\ref{Lindblad}), the corresponding matrix elements can be conveniently expressed as the difference between so-called in- and out-scattering terms (see below)
\begin{equation}\label{Lindbladalpha}
\left.\frac{d \rho_{\alpha_1\alpha_2}}{d t}\right|_{\rm scat}
=
F^{\rm in}_{\alpha_1\alpha_2}
-
F^{\rm out}_{\alpha_1\alpha_2}
\end{equation}
with
\begin{equation}\label{Fin}
F^{\rm in}_{\alpha_1\alpha_2}
=
\sum_{\alpha'_1\alpha'_2}
\mathcal{P}_{\alpha_1\alpha_2,\alpha'_1\alpha'_2}
\rho_{\alpha'_1\alpha'_2}
\end{equation}
and
\begin{equation}\label{Fout}
F^{\rm out}_{\alpha_1\alpha_2}
=
\frac{1}{2} \sum_{\alpha'_1\alpha'_2}
\mathcal{P}^*_{\alpha'_1\alpha'_1,\alpha_1\alpha'_2}
\rho_{\alpha'_2\alpha_2} \, + \textrm{H.c.}
\end{equation}
in terms of the generalized scattering rates
\begin{equation}\label{calP}
\mathcal{P}_{\alpha_1\alpha_2,\alpha'_1\alpha'_2} =
\sum_s
A^s_{\alpha_1\alpha'_1} A^{s *}_{\alpha_2\alpha'_2} \ .
\end{equation}

In order to investigate the space dependence of the phenomenon under examination ---and to compare it to its semiclassical description (see 
Sec.~\ref{s-WF} and App.~\ref{App-SL})--- let us recall the link between our density matrix $\rho_{\alpha_1\alpha_2}$ and the corresponding spatial carrier density, namely
\begin{equation}\label{n-rho}
n(\mathbf{r}) = \sum_{\alpha_1\alpha_2} \phi^{ }_{\alpha_1}(\mathbf{r}) \rho_{\alpha_1\alpha_2} \phi^*_{\alpha_2}(\mathbf{r})\ ,
\end{equation}
where $\phi_\alpha(\mathbf{r}) = \langle \mathbf{r} \vert \alpha \rangle$ denotes the real-space wavefunction  corresponding to the eigenstate $|\alpha \rangle$.
Combining the above result with the density-matrix equation (\ref{SBE}), the time evolution of the spatial carrier density is given by
\begin{equation}\label{n-tot}
\frac{\partial n(\mathbf{r})}{\partial t} 
=
\left.\frac{\partial n(\mathbf{r})}{\partial t}\right|_{\rm sp}
+
\left.\frac{\partial n(\mathbf{r})}{\partial t}\right|_{\rm scat}
\end{equation}
with
\begin{equation}\label{n-sp-rho}
\left.\frac{\partial n(\mathbf{r})}{\partial t}\right|_{\rm sp}
=
\frac{1}{i\hbar}\,\sum_{\alpha_1\alpha_2} 
\phi^{ }_{\alpha_1}(\mathbf{r}) (\epsilon_{\alpha_1}-\epsilon_{\alpha_2})\rho_{\alpha_1\alpha_2} \phi^*_{\alpha_2}(\mathbf{r})
\end{equation}
and
\begin{equation}\label{n-scat-rho}
\left.\frac{\partial n(\mathbf{r})}{\partial t}\right|_{\rm scat} =  
\sum_{\alpha_1\alpha_2} \phi^{ }_{\alpha_1}(\mathbf{r}) \Gamma(\hat{\rho})_{\alpha_1\alpha_2} \phi^*_{\alpha_2}(\mathbf{r})\ .
\end{equation}
In Sec.~\ref{s-WF} we shall show that, also for the simplest case of a bulk system, (i) in the presence of a non-parabolic band the single-particle evolution in (\ref{n-sp-rho}) deviates from the diffusion-plus drift dynamics of the semiclassical theory, and (ii) the scattering-induced variation in (\ref{n-scat-rho}) is in general different from zero, i.e., the action of the scattering superoperator is spatially non-local, in clear contrast to the Boltzmann collision term (see also App.~\ref{App-SL}).

At this point a crucial issue is in order, namely the link between the semiclassical or Boltzmann theory and the density-matrix formalism recalled so far.
As discussed in the fundamental solid-state text-book by Ashcroft and Mermin,\cite{b-Ashcroft11} a general and rigorous (i.e. quantum-mechanical) derivation of the standard semiclassical charge-transport theory constitutes a formidable task. 
The simplest approach to this tedious problem ---usually referred to as the ``diagonal limit''--- is to neglect all non-diagonal density matrix elements, which implies to assuming a single-particle density matrix of the form
\begin{equation}\label{sl1}
\rho_{\alpha_1\alpha_2} = f_{\alpha_1} \delta_{\alpha_1\alpha_2}\ .
\end{equation}
From a physical point of view, this amounts to assuming that the impact of various energy dissipation versus decoherence phenomena (described via the scattering superoperator $\Gamma$) is so strong to suppress at any time all inter-state ($\alpha_1 \neq \alpha_2$) quantum-mechanical phase coherence.
By inserting the diagonal-limit prescription (\ref{sl1}) into Eqs.~(\ref{SBE}) and (\ref{Lindbladalpha}), it is easy to get the following equation of motion for the state population $f_\alpha$:
\begin{equation}\label{sl2}
\frac{d f_\alpha}{d t} = 
\sum_{\alpha'} \left[P_{\alpha\alpha'} f_{\alpha'} - P_{\alpha'\alpha} f_\alpha\right]
\end{equation}
with
\begin{equation}\label{sl3}
P_{\alpha\alpha'} = \mathcal{P}_{\alpha\alpha,\alpha'\alpha'} = \sum_s \left|A^s_{\alpha\alpha'}\right|^2\ .
\end{equation}
Equation (\ref{sl2}) is Boltzmann-like, i.e., the time evolution of the carrier population $f_\alpha$ is dictated by a standard (in-minus-out) collision term involving scattering rates $P_{\alpha\alpha'}$ given by the diagonal elements ($\alpha_1\alpha_1' = \alpha_2\alpha_2'$) of the generalized scattering rates in (\ref{calP}).
As mentioned previously, by adopting the alternative Markov procedure proposed in Ref.~\onlinecite{Taj09b} and briefly recalled in App.~\ref{App-LSS}, for any given single-particle interaction mechanism $s$  one is able to perform a fully microscopic derivation of the corresponding Lindblad operator $\hat{A}^s$ entering the scattering superoperator (\ref{Lindblad}).
Moreover, according to this derivation, the diagonal elements of the generalized scattering rates in (\ref{sl3}) are given by the conventional Fermi's golden rule.
Indeed, the Boltzmann-like equation in (\ref{sl2}) can be regarded as the formal justification and starting point of a wide variety of Monte Carlo simulations of charge transport in semiconductor nanostructures, whose main microscopic ingredients are the carrier wavefunctions $\phi_\alpha(\mathbf{r})$ as well as the corresponding scattering rates $P_{\alpha\alpha'}$ obtained via the Fermi's golden rule.

In spite of the success of such Boltzmann-like treatment applied to the study of the steady-state electro-optical response of semiconductor nanodevices,\cite{Ryzhii98a,Iotti01b,Kohler02a,Callebaut04a,Lu06a,Bellotti08a,Jirauschek10a,Matyas10a,Iotti10a,Vitiello12a,Matyas13a,Iotti13a} the latter is not able to describe the time-dependent evolution of the spatial carrier density.
Indeed, by inserting the diagonal prescription (\ref{sl1}) into Eq.~(\ref{n-sp-rho}), the single particle contribution to the spatial carrier density is always equal to zero.
This implies that such diagonal approximation does not allow one to account for the diffusion dynamics of the semiclassical transport theory (see App.~\ref{App-SL}).
This can be easily understood noticing that within the diagonal approximation the spatial carrier density in (\ref{n-rho}) reduces to
\begin{equation}\label{n-f}
n(\mathbf{r}) = \sum_\alpha \left|\phi_\alpha(\mathbf{r})\right|^2 f_\alpha\ .
\end{equation}
This tells us that for the particular case of a bulk system ---the one considered in the conventional Boltzmann theory--- the single-particle basis states $\vert\alpha\rangle$ are momentum eigenstates, whose probability density $\left|\psi_\alpha(\mathbf{r})\right|^2$ is space-independent. 
It follows that for a bulk system the carrier density $n(\mathbf{r})$ corresponding to the above diagonal-limit picture is space-independent as well.

The obvious conclusion is that the diagonal-approximation scheme just recalled does not allow one to recover the space-dependent Boltzmann theory.
Nevertheless, as already stressed, a number of simulation strategies\cite{Ryzhii98a,Iotti01b,Kohler02a,Callebaut04a,Lu06a,Bellotti08a,Jirauschek10a,Matyas10a,Iotti10a,Vitiello12a,Matyas13a,Iotti13a,Fischetti99a,Gebauer04a,Knezevic08a} based on such diagonal-approximation paradigm came out to be quite successful in describing the steady-state electro-optical response of various semiconductor nanomaterials and devices; this is particularly true in the presence of a strong energy dissipation and decoherence, since in this case the latter dominate over scattering-free carrier diffusion (not properly described within the diagonal-approximation picture).

In order to perform a derivation of the conventional Boltzmann transport equation, it is thus vital to replace the above diagonal-approximation scheme with a genuine space-dependent description of the problem; this may be conveniently performed via the well-known Wigner picture. Indeed, during the last decades the Wigner-function formalism has been widely employed in the investigation of quantum-transport phenomena;\cite{Frensley86a,Kluksdahl89a,Buot90a,McLennan91a,Hess96a,Bordone99a,Nedjalkov04a,Querlioz08a}
however, as recently pointed out,\cite{Taj06a,Savio11a,Rosati13a,Jacoboni14a} 
such Wigner-function formalism applied to the modeling of spatially open quantum devices may lead to highly unphysical results, mainly ascribed to the failure of the conventional spatial boundary-condition scheme applied to the Wigner transport equation.
It is however imperative to stress that such limitations do not apply to the Wigner-function analysis presented below, since the latter refers to an infinitely extended system and not to a quantum device with open spatial boundaries.

\section{The Wigner-function picture and the semiclassical limit}\label{s-WF}

As anticipated, in order to account for the space-dependent character of a generic quantum nanodevice and to properly identify its semiclassical limit, a commonly employed strategy is the Wigner-function treatment of the problem.\cite{Frensley90a,b-Buot09} The Wigner function $f^{\rm W}(\mathbf{r},\mathbf{p})$ associated to a single-particle density-matrix operator $\hat{\rho}$ is defined as its Weyl-Wigner transform
\begin{equation}\label{WF-op}
f^{\rm W}(\mathbf{r},\mathbf{p}) 
= 
{\rm tr} \{ \hat{W}(\mathbf{r},\mathbf{p}) \hat{\rho} \}\ ,
\end{equation}
corresponding to the quantum-plus-statistical average of the Wigner operator\cite{b-Rossi11}
\begin{equation}\label{hatW}
\hat{W}(\mathbf{r}, \mathbf{p}) = \int d\mathbf{r}'
\left|\mathbf{r} - \frac{\mathbf{r}'}{2}\right\rangle
e^{\frac{\mathbf{p} \cdot \mathbf{r}'}{i\hbar}}
\left\langle \mathbf{r} + \frac{\mathbf{r}'}{2}\right|\ .
\end{equation}
For any physical quantity $a$ ---described via the operator $\hat{a}$--- its average value in (\ref{av1}) can be rewritten according to the Wigner picture just recalled as
\begin{equation}\label{av1-WF}
\langle a \rangle = (2\pi\hbar)^{-3}\,\int d\mathbf{r} d\mathbf{p} a^{\rm W}(\mathbf{r},\mathbf{p}) f^{\rm W}(\mathbf{r},\mathbf{p})\ ,
\end{equation}
where
\begin{equation}\label{aW}
a^{\rm W}(\mathbf{r},\mathbf{p}) = {\rm tr} \{ \hat{W}(\mathbf{r},\mathbf{p}) \hat{a} \}
\end{equation}
is the Weyl-Wigner transform of the operator $\hat{a}$.
Equation (\ref{av1-WF}) is formally identical to its semiclassical counterpart, thus confirming the central role played by the Wigner picture in establishing a direct link between the fully quantum-mechanical approach of Sec.~\ref{s-DMF} and the semiclassical Boltzmann theory (see also App.~\ref{App-SL}).
However, apart from such formal similarity, the Wigner function in (\ref{WF-op}) is not positive-definite, and cannot be regarded as a classical phase-space distribution probability.\cite{b-Kubo12a,b-Zachos05}

The time evolution of the Wigner function in (\ref{WF-op}) can be derived from the equation of motion for the density-matrix operator $\hat\rho$.
More specifically, by applying the Weyl-Wigner transform (\ref{WF-op}), together with its inverse
\begin{equation}\label{WF-op-inv}
\hat{\rho} = 
(2\pi\hbar)^{-3}\,
\int d\mathbf{r} \int d\mathbf{p} \,
\hat{W}(\mathbf{r},\mathbf{p}) \,
f^{\rm W}(\mathbf{r},\mathbf{p})  \, ,
\end{equation}
to the density-matrix equation~(\ref{DME}), one gets the equation of motion for the Wigner function:
\begin{equation}\label{WE2}
\frac{\partial f^{\rm W}(\mathbf{r},\mathbf{p})}{\partial t} = 
\left.\frac{\partial f^{\rm W}(\mathbf{r},\mathbf{p})}{\partial t}\right|_{\rm sp}
+ 
\left.\frac{\partial f^{\rm W}(\mathbf{r},\mathbf{p})}{\partial t}\right|_{\rm scat}
\end{equation}
with
\begin{equation}\label{WE2epsilon}
\left.\frac{\partial f^{\rm W}(\mathbf{r},\mathbf{p})}{\partial t}\right|_{\rm sp}
= 
\int d\mathbf{r}' \, d\mathbf{p}' \epsilon(\mathbf{r},\mathbf{p};\mathbf{r}',\mathbf{p}') f^{\rm W}(\mathbf{r}',\mathbf{p}')
\end{equation}
and
\begin{equation}\label{WE2Gamma}
\left.\frac{\partial f^{\rm W}(\mathbf{r},\mathbf{p})}{\partial t}\right|_{\rm scat}
= 
\int  d\mathbf{r}' \, d\mathbf{p}' \, \Gamma(\mathbf{r},\mathbf{p};\mathbf{r}',\mathbf{p}') f^{\rm W}(\mathbf{r}',\mathbf{p}')\ ,
\end{equation}
where 
\begin{equation}\label{epsilonWF}
\epsilon(\mathbf{r},\mathbf{p};\mathbf{r}',\mathbf{p}')
=
- \frac{i}{(2\pi)^3 \hbar^4}\,{\rm tr}\left\{
\hat{W}(\mathbf{r},\mathbf{p})\,
\left[\hat{H}_{\rm sp},\,\hat{W}(\mathbf{r}',\mathbf{p}')\right]
\right\}
\end{equation} 
and 
\begin{equation}\label{GammaWF}
\Gamma(\mathbf{r},\mathbf{p};\mathbf{r}',\mathbf{p}')
= (2\pi\hbar)^{-3}\,
{\rm tr}\left\{
\hat{W}(\mathbf{r},\mathbf{p})\,\Gamma\left(\hat{W}(\mathbf{r}',\mathbf{p}')\right)
\right\}
\end{equation} 
are the single-particle and the scattering superoperators written in the $(\mathbf{r}, \mathbf{p})$ Wigner picture, respectively.

In order to evaluate the peculiar features of the single-particle superoperator in (\ref{epsilonWF}), we shall adopt an envelope-function Hamiltonian\cite{b-Bastard88} of the form
\begin{equation}\label{EFH}
\hat{H}_{\rm sp} = K(\hat{\mathbf{p}}) + V(\hat{\mathbf{r}})\ ,
\end{equation}
where $\hat{\mathbf{r}}$ and $\hat{\mathbf{p}}$ denote, respectively, the quantum-mechanical operators associated to the electronic coordinate ($\mathbf{r}$) and momentum ($\mathbf{p}$).\footnote{According to the usual prescription of the envelope-function theory, the function $K$ in Eq.~(\ref{EFH}) describes the bulk electronic band, while $V$ describes the nanostructure  potential profile.}  
By inserting the envelope-function Hamiltonian (\ref{EFH}) into Eq.~(\ref{epsilonWF}), after a straightforward calculation (not reported here), one gets
\begin{equation}\label{WE2epsilonbis}
\left.\frac{\partial f^{\rm W}(\mathbf{r},\mathbf{p})}{\partial t}\right|_{\rm sp} = 
\left.\frac{\partial f^{\rm W}(\mathbf{r},\mathbf{p})}{\partial t}\right|_K
+ 
\left.\frac{\partial f^{\rm W}(\mathbf{r},\mathbf{p})}{\partial t}\right|_V\ ,
\end{equation}
where
\begin{equation}\label{WEK}
\left.\frac{\partial f(\mathbf{r},\mathbf{p})}{\partial t}\right|_K
= 
-\int d\mathbf{r}' \mathcal{K}(\mathbf{r}-\mathbf{r}',\mathbf{p}) f^{\rm W}(\mathbf{r}',\mathbf{p})
\end{equation}
with
\begin{equation}\label{calKbis}
\mathcal{K}(\mathbf{r}'',\mathbf{p}) \!=\! i \int\! d\mathbf{p}'
\frac{e^{-\frac{\mathbf{r}'' \cdot \mathbf{p}'}{i\hbar}}}{(2\pi)^3\hbar^4}
\!\left[K\left(\mathbf{p}\!+\!\frac{\mathbf{p}'}{2}\right)\!-\!K\left(\mathbf{p}\!-\!\frac{\mathbf{p}'}{2}\right)\right] \ ,
\end{equation}
and
\begin{equation}\label{WEV}
\left.\frac{\partial f(\mathbf{r},\mathbf{p})}{\partial t}\right|_V
= 
-\int d\mathbf{p}' \mathcal{V}(\mathbf{r},\mathbf{p}-\mathbf{p}') f^{\rm W}(\mathbf{r},\mathbf{p}')
\end{equation}
with
\begin{equation}\label{calV}
\mathcal{V}(\mathbf{r},\mathbf{p}'')  \!=\!  i \int d\mathbf{r}'
\frac{e^{\frac{\mathbf{p}'' \cdot \mathbf{r}'}{i\hbar}}}{(2\pi)^3\hbar^4}
\left[V\left(\mathbf{r}\!+\!\frac{\mathbf{r}'}{2}\right)\!-\!V\left(\mathbf{r}\!-\!\frac{\mathbf{r}'}{2}\right)\right] \ .
\end{equation}

A detailed investigation of the non-local character of the single-particle dynamics in (\ref{WE2epsilonbis}) --- induced by the kinetic superoperator $\mathcal{K}$ in (\ref{WEK}) as well as by the potential superoperator $\mathcal{V}$ in (\ref{WEV})--- 
can be found in Ref.~\onlinecite{Rosati13a}.

Let us now discuss the general non-local features of the scattering superoperator in (\ref{GammaWF}).
By inserting into Eq.~(\ref{GammaWF}) the explicit form of the Lindblad-like superoperator (\ref{Lindblad}), we get:
\begin{widetext}
\begin{equation}\label{GammaWFbis}
\Gamma(\mathbf{r},\mathbf{p};\mathbf{r}',\mathbf{p}')
= 
(2\pi\hbar)^{-3}\,
\sum_s 
\Re\left\{
{\rm tr}\left\{
\hat{W}(\mathbf{r},\mathbf{p})\,
\hat{A}^s
\hat{W}(\mathbf{r}',\mathbf{p}')
\hat{A}^{\dagger s}
\right\} 
- 
{\rm tr}\left\{
\hat{W}(\mathbf{r},\mathbf{p})\,
\hat{A}^{s \dagger} \hat{A}^s
\hat{W}(\mathbf{r}',\mathbf{p}')
\right\}\right\} 
\ .
\end{equation} 
\end{widetext}
As shown below (see also App.~\ref{App-SL}), in the so-called semiclassical limit these two contributions reduce to the in- and out-scattering terms of the Boltzmann theory (see Eq.~(\ref{bte4WF})); however, opposite to the Boltzmann collision term, the quantum-mechanical scattering superoperator in (\ref{GammaWFbis}) is in general spatially non-local.
Indeed, for a generic Lindblad operator $\hat{A}^s$ corresponding to a given interaction mechanism $s$, the scattering superoperator is different from zero also for $\mathbf{r} \neq \mathbf{r}'$.

In order to better elucidate the spatial nonlocality of the Wigner-transport theory,
it is useful to recall the link between our Wigner function $f^{\rm W}(\mathbf{r},\mathbf{p})$ and the corresponding spatial carrier density $n(\mathbf{r})$; according to the general average-value prescription (\ref{av1-WF}), one gets a result formally identical to the semiclassical one, namely:
\begin{equation}\label{nWF}
n(\mathbf{r}) = (2\pi\hbar)^{-3}\,\int d^3p\, f^{\rm W}(\mathbf{r},\mathbf{p})\ .
\end{equation}
Combining the above result with the Wigner transport equation (\ref{WE2}) and employing the single-particle results in (\ref{WE2epsilonbis})-(\ref{calV}), the time evolution of the spatial carrier density is again given by Eq.~(\ref{n-tot}) with
\begin{equation}\label{n-sp-WF}
\left.\frac{\partial n(\mathbf{r})}{\partial t}\right|_{\rm sp}
=
- (2\pi\hbar)^{-3} \int d\mathbf{r}' d\mathbf{p}' \mathcal{K}(\mathbf{r}-\mathbf{r}',\mathbf{p}') f^{\rm W}(\mathbf{r}',\mathbf{p}')
\end{equation}
and
\begin{equation}\label{n-scat-WF}
\left.\frac{\partial n(\mathbf{r})}{\partial t}\right|_{\rm scat}
\!=\! 
(2\pi\hbar)^{-3}
\int  d\mathbf{r}' d\mathbf{p} d\mathbf{p}'  
\Gamma(\mathbf{r}\!,\!\mathbf{p};\mathbf{r}'\!,\!\mathbf{p}') f^{\rm W}(\mathbf{r}'\!,\!\mathbf{p}')\ .
\end{equation}
It is important to stress that, also within the present quantum-mechanical treatment, the time evolution of the spatial carrier density in Eq.~(\ref{n-tot}) can be expressed via the usual charge continuity equation, i.e.,
\begin{equation}\label{n-cce}
\frac{\partial n(\mathbf{r})}{\partial t}
+ \mathbf{\nabla} \cdot \mathbf{J}(\mathbf{r}) = 0\ .
\end{equation}
To this end, the carrier current density $\mathbf{J}(\mathbf{r})$ is defined as the average value (see Eqs.~(\ref{av1-WF}) and (\ref{aW})) of a corresponding quantum-mechanical operator $\hat{\mathbf{J}}(\mathbf{r})$ as
\begin{equation}\label{J-WF}
\mathbf{J}(\mathbf{r}) = (2\pi\hbar)^{-3}\,\int d\mathbf{r}' d\mathbf{p}' \mathbf{J}^{\rm W}(\mathbf{r}; \mathbf{r}',\mathbf{p}') f^{\rm W}(\mathbf{r}',\mathbf{p}')\ ,
\end{equation}
where
\begin{equation}\label{JW}
\mathbf{J}^{\rm W}(\mathbf{r}; \mathbf{r}',\mathbf{p}') 
= 
{\rm tr} \{ \hat{W}(\mathbf{r}',\mathbf{p}') \hat{\mathbf{J}}(\mathbf{r})\}
\end{equation}
is the Weyl-Wigner transform of the current-density operator. 
Combining Eqs.~(\ref{n-tot}), (\ref{n-cce}), (\ref{n-sp-WF}), and (\ref{n-scat-WF}), after a straightforward calculation (not reported here) one gets
\begin{equation}\label{JWbis}
\mathbf{J}^{\rm W}(\mathbf{r}; \mathbf{r}',\mathbf{p}') 
=
\mathbf{J}^{\rm W}_{\rm sp}(\mathbf{r}; \mathbf{r}',\mathbf{p}') 
+
\mathbf{J}^{\rm W}_{\rm scat}(\mathbf{r}; \mathbf{r}',\mathbf{p}')
\end{equation}
with
\begin{equation}\label{JW-sp}
\mathbf{J}^{\rm W}_{\rm sp}(\mathbf{r}; \mathbf{r}',\mathbf{p}') 
\!=\!
(2\pi\hbar)^{-3} 
\int d\mathbf{r}'' d\mathbf{p}'' 
\frac{e^{\frac{\mathbf{p}'' \cdot (\mathbf{r}''-\mathbf{r})}{i\hbar}}}{i \mathbf{p}''}
\mathcal{K}(\mathbf{r}''-\mathbf{r}',\mathbf{p}') 
\end{equation}
and
\begin{widetext}
\begin{equation}\label{JW-scat}
\mathbf{J}^{\rm W}_{\rm scat}(\mathbf{r}; \mathbf{r}',\mathbf{p}') 
=
-(2\pi\hbar)^{-3} 
\int d\mathbf{r}'' d\mathbf{p} d\mathbf{p}'' 
\frac{e^{\frac{\mathbf{p}'' \cdot (\mathbf{r}''-\mathbf{r})}{i\hbar}}}{i \mathbf{p}''}
\Gamma(\mathbf{r}'',\mathbf{p};\mathbf{r}',\mathbf{p}') 
\ .
\end{equation}
\end{widetext}
It follows that the quantum-mechanical current density in (\ref{J-WF}) is the sum of a single-particle and of a scattering contribution; it is worth stressing that the presence of a scattering-induced current has been clearly pointed out by Gebauer and Car in Ref.~\onlinecite{Gebauer04a}.

While for the particular case of a parabolic band the kinetic term of the Wigner equation reduces to the diffusion term of the Boltzmann theory (see below) and the single-particle current is simply given by 
\begin{equation}\label{J-WF-bis}
\mathbf{J}_{\rm sp}(\mathbf{r}) = (2\pi\hbar)^{-3}\,\int d^3p\, \mathbf{v}(\mathbf{p}) \, f^{\rm W}(\mathbf{r},\mathbf{p})\ ,
\end{equation}
for non-parabolic bands the single-particle current density is always described in terms of the spatially non-local superoperator in (\ref{JW-sp}).\cite{Hess96a,Demeio05a,Rosati13a}

The explicit form of the scattering-induced current-density operator in (\ref{JW-scat}) will depend strongly on the specific form of the scattering superoperator $\Gamma$. In any case, opposite to the semiclassical scenario, within a fully quantum-mechanical description such scattering-induced current is in general different from zero, which is again a clear fingerprint of the non-local character of our scattering superoperator.

Let us finally discuss the so-called semiclassical limit, namely how to recover the Boltzmann transport equation as the limit of the above Wigner transport theory for $\hbar \to 0$.

As far as the single-particle contribution in (\ref{WE2epsilonbis}) is concerned, this limit is well established, and can be straightforwardly performed expressing such single-particle dynamics in terms of the well-known Moyal 
brackets;\cite{Moyal49a} indeed, for $\hbar \to 0$ the latter reduce to the usual Poisson brackets of classical mechanics, which in our case correspond to the usual diffusion-plus-drift terms of the Boltzmann theory.

The most difficult task of the semiclassical limit is to show that for $\hbar \to 0$ the (spatially non-local) scattering superoperator in (\ref{WE2Gamma}) reduces to the (spatially local) collision term of the Boltzmann theory.
Indeed, as shown in App.~\ref{App-SL}, by employing the momentum representation and applying an adiabatic-decoupling scheme (valid for $\hbar \to 0$) both in the coordinate and momentum space, one finally gets 
\begin{equation}\label{bte4WF}
\left.\frac{\partial f^{\rm W}}{\partial t}\right|_{\rm scat} \!=\!
\int d^3p'
\left[
P(\mathbf{p},\mathbf{p}') f^{\rm W}(\mathbf{r}\!,\!\mathbf{p}')
\!-\!
P(\mathbf{p}'\!,\!\mathbf{p}) f^{\rm W}(\mathbf{r}\!,\!\mathbf{p})\right],
\end{equation}
where the semiclassical scattering rates $P(\mathbf{p},\mathbf{p}')$ can easily be expressed in terms of the matrix elements of the original Lindblad operators (see Eqs.~(\ref{P}) and (\ref{PsWF}) in App.~\ref{App-SL}).

\section{Scattering-induced diffusion: A few simulated experiments}\label{s-sid}

Aim of this section is to perform a detailed investigation of scattering-induced diffusion in homogeneous as well as in nanostructured semiconductor systems. Based on the quantum-transport formulation proposed so far, we shall present and discuss a number of simulated experiments of ultrafast carrier dynamics in GaN-based materials.

\subsection{Physical model and simulation strategy}\label{ss-psss}

As prototypical physical system we shall consider an effective one-dimensional GaN-based nanostructure, whose main energy-dissipation and decoherence mechanism is carrier-LO phonon scattering. The latter will be described via the Lindblad scattering superoperator in (\ref{Lindblad}), whose explicit form is given in App.~\ref{App-LSS}.

It is imperative to stress that the choice of considering a simple one-dimensional model is by no means dictated by computational limits; indeed, opposite to more refined quantum-kinetic approaches, the proposed simulation strategy may be easily applied to realistic nanostructures within a fully three-dimensional description, as recently realized in Ref.~\onlinecite{Dolcini13a}.
We just decided to adopt a one-dimensional system in order to facilitate the analysis of scattering-induced spatial nonlocality, and to better elucidate its physical origin and magnitude.

For the case of a one-dimensional system with coordinate $z$ and momentum $p$, the space (see Eq.~(\ref{n-rho})) and momentum charge distributions are simply given by
\begin{equation}\label{nz}
n(z) = \sum_{\alpha_1\alpha_2} \phi^{ }_{\alpha_1}(z) \rho_{\alpha_1\alpha_2} \phi^*_{\alpha_2}(z)
\end{equation}
and
\begin{equation}\label{nk}
n(p) = \sum_{\alpha_1\alpha_2} \tilde{\phi}^{ }_{\alpha_1}(p) \rho_{\alpha_1\alpha_2} \tilde{\phi}^*_{\alpha_2}(p)\ ,
\end{equation}
where $\phi_\alpha(z) \equiv \langle z \vert \alpha \rangle$ denotes the real-space wavefunction  corresponding to the eigenstate $\alpha$, and $\tilde{\phi}_\alpha(p) \equiv \langle p \vert \alpha \rangle$ its Fourier transform.

Combining the prescription in (\ref{nz}) with the density-matrix equation (\ref{SBE}), the total time evolution of the spatial carrier density $n(z)$ is described via the one-dimensional versions ($\mathbf{r} \to z$) of Eqs.~(\ref{n-tot})-(\ref{n-scat-rho}).
As already pointed out in Sec.~\ref{s-WF}, for the relevant case of the Lindblad superoperator in (\ref{Lindbladalpha}) the corresponding time evolution can be expressed as the difference of two terms, which in the semiclassical limit (see also App.~\ref{App-SL}) reduce to the in- minus out-scattering structure of the conventional Boltzmann theory (see Eq.~(\ref{bte4WF})). This suggests to write the one-dimensional version of Eq.~(\ref{n-scat-rho}) as
\begin{equation}\label{dndt-scat}
\left.\frac{\partial n(z)}{\partial t}\right|_{\rm scat} 
=
F^{\rm in}(z)
-
F^{\rm out}(z)
\end{equation} 
with
\begin{equation}\label{dndt-inout}
F^{\rm in/out}(z)
= 
\sum_{\alpha_1\alpha_2} 
\phi^{ }_{\alpha_1}(z) 
F^{\rm in/out}_{\alpha_1\alpha_2}
\phi^*_{\alpha_2}(z)\ .
\end{equation}

Our simulation strategy is based on a numerical solution of the density-matrix equation in (\ref{SBE}); this is realized via a fixed-time-step discretization\cite{b-Rossi11} based on an exact integration of the single-particle dynamics.
More specifically, the single-particle states $\alpha$ of the structure under examination are described via the usual envelope-function picture (see Eq.~(\ref{EFH})) within the standard effective-mass approximation,\cite{b-Bastard88} in terms of a plane-wave expansion.\cite{b-Rossi11} 

In order to mimic the main features of a realistic GaN-based material, the following parameters have been employed:
effective mass $m^* = 0.2 m_\circ$ ($m_\circ$ denoting the free-electron one) and LO-phonon energy $\epsilon_{\rm LO} = 80$\,meV; 
moreover, the amplitude of the carrier-phonon matrix elements in Eq.~(\ref{Hprimecp}) are chosen such to reproduce an average bulk carrier-LO phonon scattering rate $\tau_{\rm LO} = 25$\,fs. 

For all the simulated experiments presented below we have chosen as initial condition a single-particle density matrix $\overline{\rho}_{\alpha_1\alpha_2}$ corresponding to a gaussian carrier distribution both in space and momentum, namely
\begin{equation}\label{Gauss}
\overline{n}(z) \propto \frac{e^{-\frac{z^2}{2 \overline{\Delta}_z^2}}}{\sqrt{2\pi} \,\overline{\Delta}_z}
\ , \qquad
\overline{n}(p) \propto \frac{e^{-\frac{p^2}{2 \overline{\Delta}_p^2}}}{\sqrt{2\pi}\, \overline{\Delta}_p}\ ,
\end{equation}
where $\overline{\Delta}_z$ describes the degree of spatial localization of our initial state, and $\overline{\Delta}_p = \sqrt{m^*k_B T}$ 
describes the thermal fluctuations of our carrier gas.

It is easy to show that such initial condition corresponds to a one-dimensional Wigner function
\begin{equation}\label{fWcirc}
\overline{f}^{\rm W}(z,p) \propto \hbar\,
\frac{
e^{-\frac{z^2}{2 \overline{\Delta}_z^2}}
e^{-\frac{p^2}{2 \overline{\Delta}_p^2}}
}
{
\sqrt{2\pi}\, \overline{\Delta}_z \overline{\Delta}_p
}
\ ,
\end{equation}
and therefore to an initial density matrix
\begin{equation}\label{rhocirc}
\overline{\rho}_{\alpha_1\alpha_2} \propto 
\frac{1}{2\pi}\,\int dz\, dp 
W_{\alpha_1\alpha_2}(z,p)
\frac{
e^{-\frac{z^2}{2 \overline{\Delta}_z^2}}
e^{-\frac{p^2}{2 \overline{\Delta}_p^2}}
}
{
\sqrt{2\pi}\, \overline{\Delta}_z \overline{\Delta}_p
}
\ ,
\end{equation}
where
\begin{equation}\label{calW}
W_{\alpha_1\alpha_2}(z,p) = \int dz'
\phi^*_{\alpha_1}\left(z-\frac{z'}{2}\right)
e^{\frac{p z'}{i\hbar}}
\phi^{ }_{\alpha_2}\left(z+\frac{z'}{2}\right) 
\end{equation}
are the single-particle matrix elements of the Wigner operator in (\ref{hatW}).\footnote{We stress that the (mixed-state) density matrix in (\ref{rhocirc}) is not always physical; indeed, it is possible to show that the uncertainty principle imposes the following restriction:
$\overline{\Delta}_z \ge \frac{\hbar}{2 \overline{\Delta}_p}$. 
Recalling that $\overline{\Delta}_p = \sqrt{m^*k_B T}$, it follows that at room temperature and for the GaN parameters previously recalled, one gets: 
$\overline{\Delta}_z \ge \frac{\hbar}{2\sqrt{m^*k_B T}} \simeq 2$\,nm.}

Primary goal of our simulated experiments is to investigate the non-local character of the Lindblad-like scattering superoperator in (\ref{Lindbladalpha}), and to compare it with other scattering models.
The simplest parameter-free form of the scattering term entering our density-matrix equation (\ref{SBE}) is given by the following relaxation-time model:\cite{Dolcini13a}
\begin{equation}\label{RTA}
\left.\frac{d \rho_{\alpha_1\alpha_2}}{dt}\right|_{\rm scat}
=
-\,\frac{\Gamma_{\alpha_1} + \Gamma_{\alpha_2}}{2}
\left(\rho_{\alpha_1\alpha_2} - \rho^\circ_{\alpha_1\alpha_2}\right)\ .
\end{equation}
Here $\rho^\circ_{\alpha_1\alpha_2} = f^\circ_{\alpha_1} \delta_{\alpha_1\alpha_2}$ is the equilibrium density matrix dictated by the host material, and
\begin{equation}\label{Gamma}
\Gamma_\alpha = \sum_s \sum_{\alpha'} P^s_{\alpha'\alpha}
\end{equation}
is the total scattering rate (i.e., summed over all final states $\alpha'$ and relevant interaction mechanisms $s$) corresponding to the microscopic transition probabilities $P^s_{\alpha'\alpha}$ of the semiclassical transport theory given by the standard Fermi's golden rule.\cite{b-Jacoboni89}
Within such relaxation-time paradigm, the diagonal contributions ($\alpha_1 = \alpha_2$) describe population transfer (and thus energy dissipation) toward the equilibrium carrier distribution $f^\circ_{\alpha_1}$ according to the relaxation rate $\Gamma_{\alpha_1}$, whereas the off-diagonal contributions ($\alpha_1 \ne \alpha_2$) describe a decay of the inter-state polarizations according to the decoherence rate 
$(\Gamma_{\alpha_1} + \Gamma_{\alpha_2})/2$.

In spite of its simple form and straightforward physical interpretation, the structure of the relaxation-time term (\ref{RTA}) is intrinsically different from the in- minus out-structure of the Boltzmann collision term as well as of the Lindblad superoperator in (\ref{Lindbladalpha}), and for this reason it may lead to a significant overestimation of decoherence processes (see below). 

\subsection{Analysis of homogeneous systems}\label{ss-abs}

Our first set of room-temperature simulated experiments corresponds to an effective (one-dimensional) homogeneous GaN system (i.e., no confinement potential profile along the $z$ direction). 

\subsubsection{Scattering nonlocality}\label{sss-snl}

\begin{figure}
\centering
\includegraphics*[width=7.2cm]{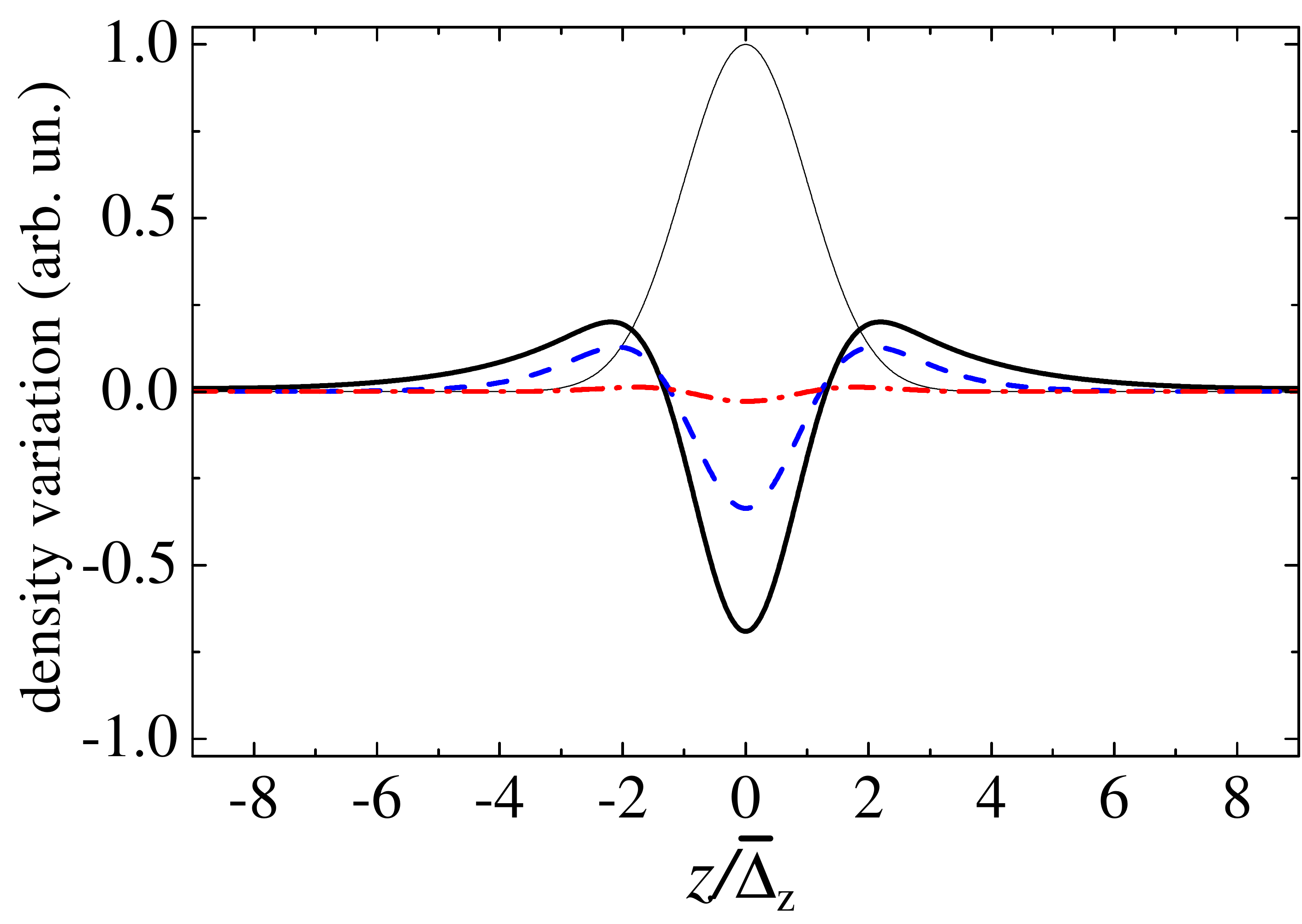}
\caption[]{(Color online)
Room-temperature carrier-LO phonon scattering nonlocality induced by the Lindblad superoperator in Eq.~(\ref{Lindbladalpha}) in a homogeneous GaN system: scattering-induced time derivative of the spatial carrier density (see Eq.~(\ref{dndt-scat})) as a function of the relative coordinate $z/\overline{\Delta}_z$ for three different values of the localization parameter: $\overline{\Delta}_z = 5$\,nm (solid curve), $\overline{\Delta}_z = 10$\,nm (dashed curve), and $\overline{\Delta}_z = 50$\,nm (dash-dotted curve), together with the initial spatial density profile in Eq.~(\ref{Gauss}) (thin solid curve) (see text).
}
\label{Fig1}       
\end{figure}

Let us start our analysis by investigating the carrier-LO phonon scattering nonlocality induced by the Lindblad superoperator in (\ref{Lindbladalpha}).
Figure \ref{Fig1} shows the scattering-induced time derivative of the spatial carrier density (see Eq.~(\ref{dndt-scat})) as a function of the relative coordinate $z/\overline{\Delta}_z$ for three different values of the localization parameter $\overline{\Delta}_z$. 
As we can see, in the presence of an initial nanometric confinement (solid and dashed curves) the phonon-induced time variation is significantly different from zero; the latter displays a negative peak ---corresponding to a sort of replica of the initial distribution--- and, more importantly, a positive contribution extending over a much larger range.
This is exactly the signature of scattering-induced spatial nonlocality we were looking for.
By significantly increasing the value of $\overline{\Delta}_z$ (dash-dotted curve), the magnitude and relative spatial extension of such nonlocality effects is strongly reduced, thus confirming that in the semiclassical limit $\overline{\Delta}_z \to \infty$ the scattering-induced time variation tends to zero, as predicted by the conventional Boltzmann theory (see App.~\ref{App-SL}).

\begin{figure}
\centering
\includegraphics*[width=7.2cm]{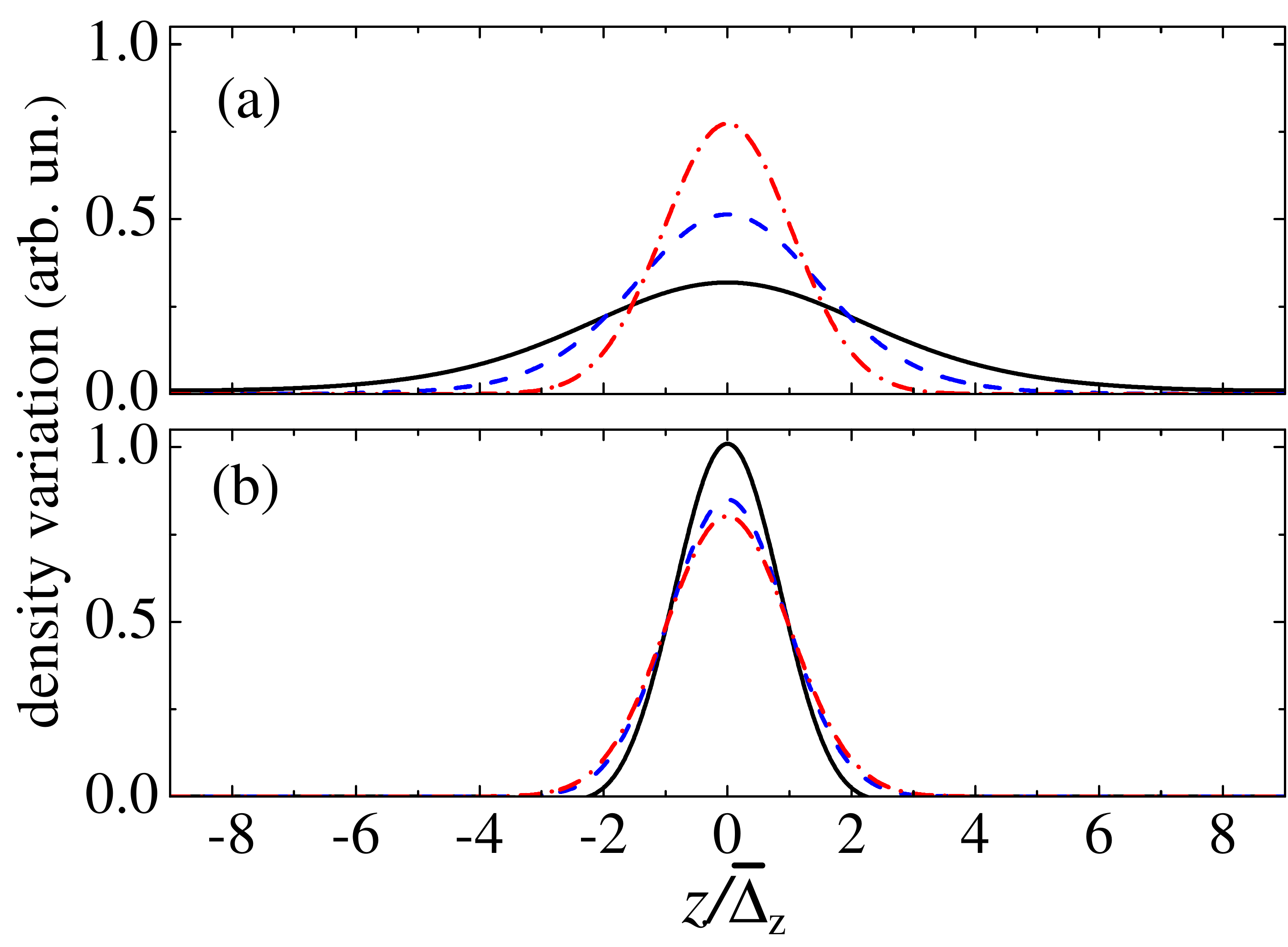}
\caption[]{(Color online)
Room-temperature carrier-LO phonon scattering nonlocality induced by the Lindblad superoperator in Eq.~(\ref{Lindbladalpha}) in a homogeneous GaN system:
in- (panel a) and out-scattering contributions (panel b) corresponding to the time derivatives of the spatial carrier density (see Eq.~(\ref{dndt-scat})) reported in Fig.~\ref{Fig1} (see text).
}
\label{Fig2}       
\end{figure}

In order to better understand the physical origin and relative magnitude of the positive versus negative regions in Fig.~\ref{Fig1}, let us examine separately the impact of in- and out-scattering terms (see Eq.~(\ref{dndt-scat})).
Figure \ref{Fig2} shows in- (panel a) and out-scattering contributions (panel b) corresponding to the time derivatives of the spatial carrier density (see Eq.~(\ref{dndt-scat})) reported in Fig.~\ref{Fig1}.
As we can see, in the presence of an initial nanometric confinement (solid and dashed curves) the in-scattering contribution (panel a) is significantly larger than the initial distribution profile (see thin solid curve in Fig.~\ref{Fig1}) while, in contrast, the out-scattering contribution (panel b) comes out to be more localized.
It is exactly such different spatial extension of in- and out-scattering contributions that gives rise to the density-variation profiles in Fig.~\ref{Fig1}; in particular, the significant delocalization of the in-scattering contribution (compared to the out-scattering one) is responsible (i) of the negative central peak, and (ii) of the two positive external regions.\footnote{It is worth stressing that, in view of the trace-preserving character of the Lindblad superoperator (\ref{Lindblad}), the total carrier density (i.e., integrated over the spatial coordinate $z$) is preserved; this implies that the positive and negative regions in Fig.~\ref{Fig1} should cancel each other out.}
By significantly increasing the value of $\overline{\Delta}_z$ (dash-dotted curves), in- and out-scattering contributions tend to coincide, which implies that their difference tends to vanish, in total agreement with the corresponding result in Fig.~\ref{Fig1} (dash-dotted curve). This clearly shows that the local character of the Boltzmann theory (see Eq.~(\ref{bte4WF})) originates from an exact cancelation between in- and out-scattering contributions, which takes place in the semiclassical limit (i.e., $\overline{\Delta}_z \to \infty$) only.

Based on the numerical results presented so far, it is easy to conclude that the impact of scattering nonlocality is intimately related to the different spatial extension of in- and out-scattering contributions.
In order to better quantify the phenomenon under examination, it is useful to introduce the effective nonlocality parameter
\begin{equation}\label{eta}
\eta^{\rm in/out}
 = 
\frac{1}{\overline{\Delta}_z}\,
\sqrt{
\frac{
\int z^2 \left| F^{\rm in/out}(z) \right| dz
}{
\int \left| F^{\rm in/out}(z) \right| dz
}
}\ .
\end{equation}
According to its definition, this dimensionless parameter can be regarded as the standard deviation of the spatial density variation $F^{\rm in/out}(z)$ (see Eq.~(\ref{dndt-scat})) in units of $\overline{\Delta}_z$.
It follows that when the shape of the density variation $F^{\rm in/out}(z)$ tends to the initial Gaussian profile (see dash-dotted curves in Fig.~\ref{Fig2}), the nonlocality parameter $\eta^{\rm in/out}$ in (\ref{eta}) tends to one; moreover, for charge variations wider than the initial distribution (see solid and dashed curves in Fig.~\ref{Fig2}a) the nonlocality parameter is expected to be greater than one, while for charge variations sharper than the initial distribution (see solid and dashed curves in Fig.~\ref{Fig2}b) the latter is expected to be smaller than one.

\begin{figure}
\centering
\includegraphics*[width=7.2cm]{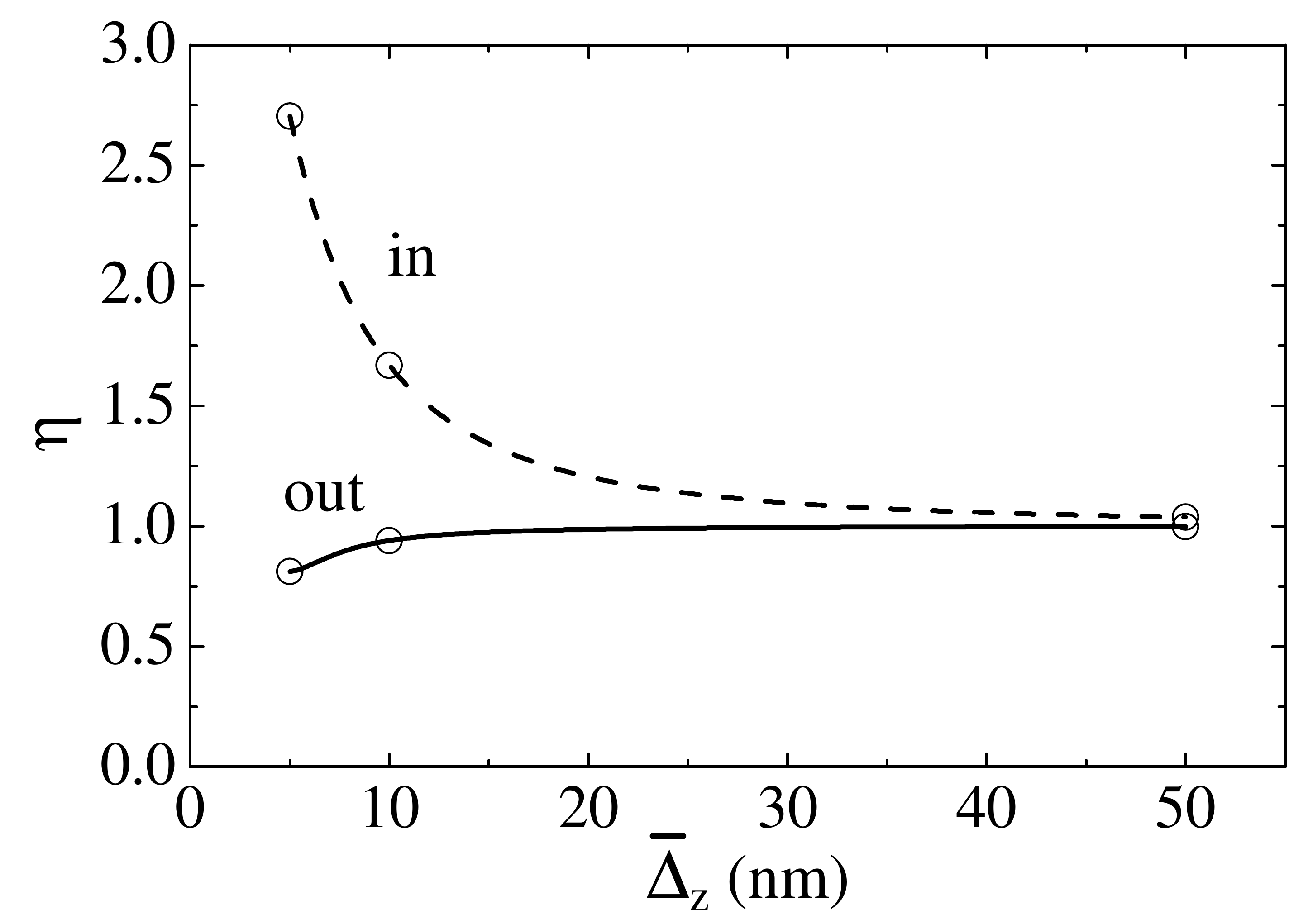}
\caption[]{(Color online)
Nonlocality parameter in Eq.~(\ref{eta}) as a function of $\overline{\Delta}_z$ for both in- and out-scattering contributions.
Here, the $5$\,nm, $10$\,nm, and $50$\,nm values (see symbols) correspond to the in- and out-scattering profiles of Fig.~\ref{Fig2} (see text).
}
\label{Fig3}       
\end{figure}

This scenario is fully confirmed by the numerical results reported in Fig.~\ref{Fig3}, where the nonlocality parameter in (\ref{eta}) is plotted as a function of $\overline{\Delta}_z$ for both in- and out-scattering contributions (here, the two curves have been obtained repeating our numerical calculation for a large set of $\overline{\Delta}_z$ values).
As we can see, in the presence of a strong spatial confinement ($\overline{\Delta}_z = 5$\,nm) (see solid curves in Fig.~\ref{Fig2}) the nonlocality parameter of the in-scattering term is definitely greater than one, while for the out-scattering term the latter is significantly smaller than one.
By increasing the value of $\overline{\Delta}_z$, the difference between in- and out-parameters is progressively reduced, and for $\overline{\Delta}_z = 50$\,nm (see dash-dotted curves in Fig.~\ref{Fig2}) their value is already very close to unity.

The homogeneous-GaN simulated experiments presented so far allows one to draw two basic conclusions: (i) in the presence of a nanometric spatial confinement one deals with a significant carrier-phonon scattering nonlocality (see solid curve in Fig.~\ref{Fig1}); (ii) opposite to other simplified scattering models (see below), our Lindblad superoperator (see Eq.~(\ref{Lindbladalpha})) is able to properly reproduce the semiclassical-limit behavior (see dash-dotted curve in Fig.~\ref{Fig1}), thus recovering the local character of the Boltzmann collision term.

\begin{figure}
\centering
\includegraphics*[width=7.2cm]{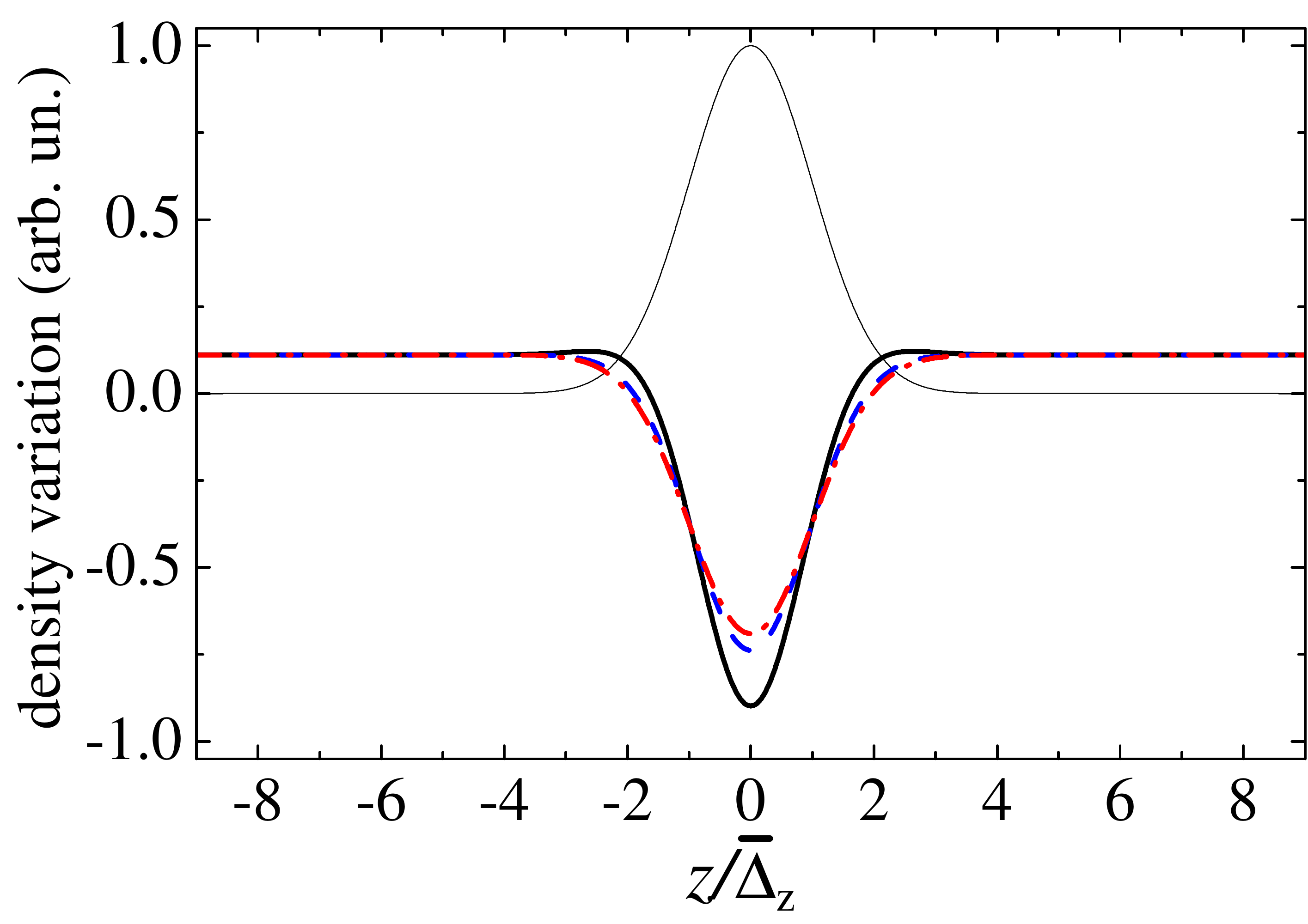}
\caption[]{(Color online)
Same as in Fig.~\ref{Fig1} but for the relaxation-time model in Eq.~(\ref{RTA}) (see text).
}
\label{Fig4}       
\end{figure}

At this point it is crucial to compare the action of the Lindblad scattering superoperator (\ref{Lindbladalpha}) (see Fig.~\ref{Fig1}) with that of simplified dissipation models, and in particular with the conventional relaxation-time approximation.
Figure \ref{Fig4} shows the scattering-induced time derivative of the spatial carrier density corresponding to the relaxation-time model in (\ref{RTA}) as a function of the relative coordinate $z/\overline{\Delta}_z$ for the same three values of the localization parameter $\overline{\Delta}_z$ considered in Fig.~\ref{Fig1}.
As we can see, also for the case of the relaxation-time model one deals with significant nonlocality effects. However, comparing Fig.~\ref{Fig4} with Fig.~\ref{Fig1}, it is easy to recognize strong differences between the Lindblad treatment and the relaxation-time approximation:
opposite to the Lindblad-superoperator results of Fig.~\ref{Fig1}, here the shape and amplitude of the charge-density variation is not strongly influenced by the value of $\overline{\Delta}_z$; 
more importantly, while in Fig.~\ref{Fig1} the positive regions are spatially localized (i.e., they display a maximum and then vanish at large distances), here the charge variation tends to a constant and $\overline{\Delta}_z$-independent value.
This constitutes an unambiguous proof of the intrinsic limitations of the relaxation-time approximation; indeed, opposite to the Lindblad-superoperator treatment, the latter (i) comes out to be totally non-local (as confirmed by its nearly constant values at large coordinate values),\footnote{Indeed, for the relaxation-time model in (\ref{RTA}) it is not possible to introduce a nonlocality parameter (see Eq.~(\ref{eta})), since the spatial standard deviation of the charge-density variation in Fig.~\ref{Fig4} is always infinite.}
and (ii) in the semiclassical limit ($\overline{\Delta}_z \to \infty$) it is intrinsically unable to reproduce the local character of the Boltzmann collision term.

As we shall see, the totally non-local character of the relaxation-time model may give rise to a strong overestimation of the scattering-induced quantum diffusion (see Figs.~\ref{Fig6} and \ref{Fig7} below).

\subsubsection{Quantum diffusion: single-particle versus scattering dynamics}\label{sss-aqd}

So far our focus has been devoted to the investigation of the spatial nonlocality induced by carrier-LO phonon coupling. However, in order to establish how such scattering-induced charge redistribution will affect the overall diffusion process, it is imperative to perform a time-dependent analysis including single-particle as well as scattering dynamics.

\begin{figure}
\centering
\includegraphics*[width=7.2cm]{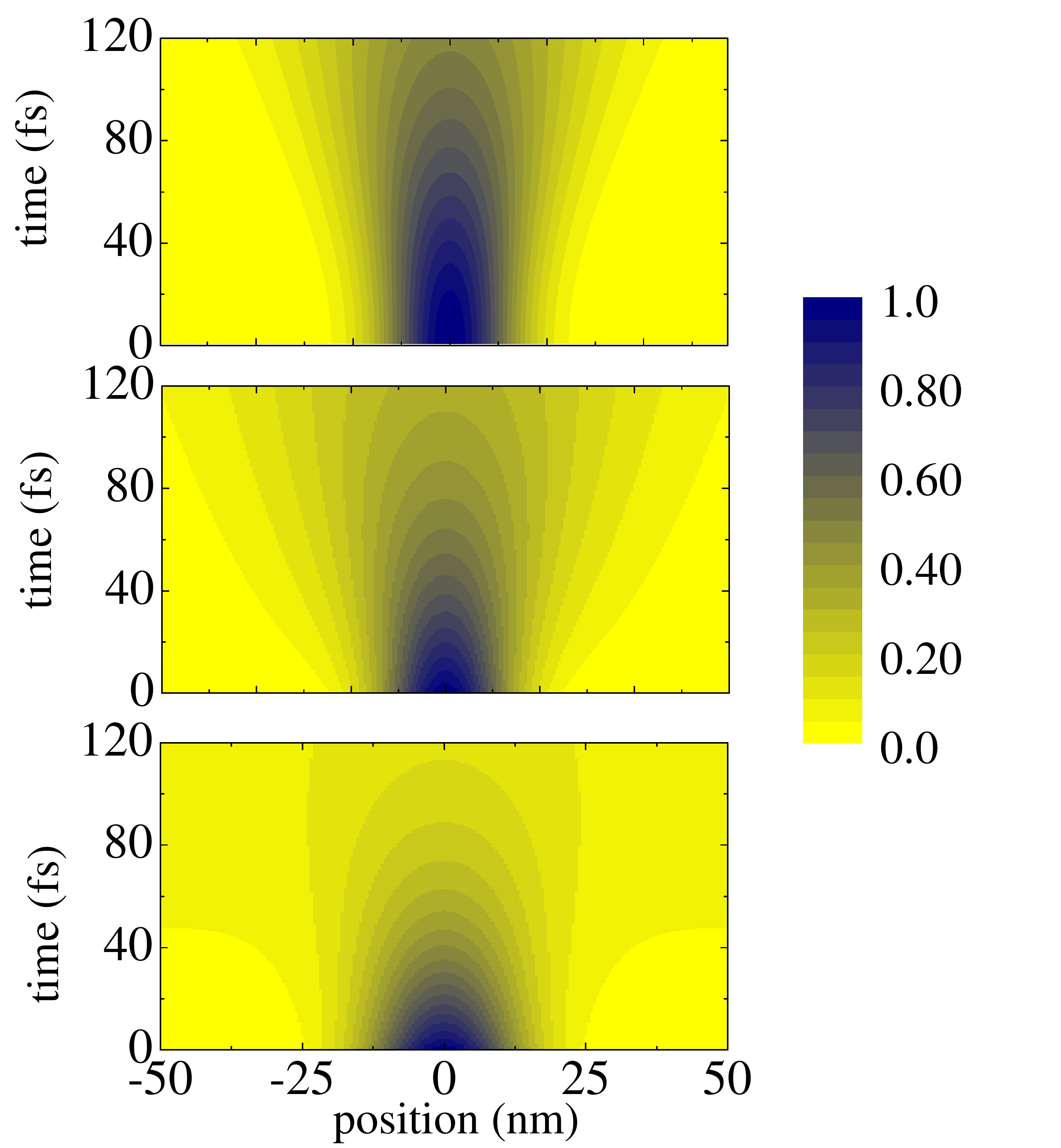}
\caption[]{(Color online)
Room-temperature quantum-diffusion dynamics in a homogeneous GaN system obtained in the absence of carrier-phonon coupling (upper panel), via the Lindblad scattering superoperator in Eq.~(\ref{Lindbladalpha}) (central panel), and via the relaxation-time model in Eq.~(\ref{RTA}) (lower panel): sub-picosecond time evolution of the spatial carrier density corresponding to the initial mixed state in Eq.~(\ref{rhocirc}) with $\overline{\Delta}_z = 10$\,nm (see text).
}
\label{Fig5}
\end{figure}

Figure \ref{Fig5} displays the sub-picosecond time evolution of the spatial carrier density corresponding to the initial mixed state in (\ref{rhocirc}) with $\overline{\Delta}_z = 10$\,nm, obtained in the absence of carrier-phonon coupling (upper panel), via the Lindblad scattering superoperator in (\ref{Lindbladalpha}) (central panel), and via the relaxation-time model in (\ref{RTA}) (lower panel).
As we can see, compared to the scattering-free case (upper panel), both Lindblad and relaxation-time treatments give rise to a speed up of the diffusion process, and the effect is more pronounced in the relaxation-time case (lower panel).

Such ultrafast diffusion dynamics is the result of a highly non-trivial interplay between single-particle and scattering contributions; indeed, it is well known that also in the presence of a spatially local (i.e., Boltzmann) scattering model (for which the contribution in (\ref{dndt-scat}) is always equal to zero) any  scattering-induced carrier redistribution tends to speed up the diffusion process.\cite{b-Ashcroft11}
In order to better evaluate the genuine diffusion contribution due to scattering nonlocality, it is then crucial to start our simulated experiments from a thermalized carrier distribution; this has been realized adopting the initial state in (\ref{rhocirc});
indeed, for a parabolic-band homogeneous system (as the one considered here) in the absence of scattering nonlocality, the time evolution of the spatial carrier density is described by the following (time-dependent) Gaussian distribution (see upper panel in Fig.~\ref{Fig5})
\begin{equation}\label{nzt}
n(z,t) \propto \frac{e^{-\frac{z^2}{2 \Delta_z^2(t)}}}{\sqrt{2\pi} \Delta_z(t)}
\end{equation}
with
\begin{equation}\label{Deltazt}
\Delta_z(t) = \overline{\Delta}_z \sqrt{1+\frac{t^2}{\tau_d^2}}\ ,
\end{equation}
where 
\begin{equation}\label{taud}
\tau_d = \frac{m^* \overline{\Delta}_z}{\overline{\Delta}_p} 
\end{equation}
describes the typical time scale of the scattering-free diffusion process (for the case of Fig.~\ref{Fig5} this is about $70$\,fs).

The physical origin and relative magnitude of the diffusion speed up reported in Fig.~\ref{Fig5} can be easily understood in terms of the scattering-induced nonlocality previously investigated.
Indeed, for both the Lindblad (Fig.~\ref{Fig1}) and the relaxation-time model (Fig.~\ref{Fig4}), carrier-phonon scattering induces a progressive charge transfer from the initial peak toward outer regions, which results in an overall spatial broadening.
As already pointed out, the impact of such scattering-induced diffusion is expected to be particularly pronounced in the case of the relaxation-time model, since the latter is totally non-local (see Fig.~\ref{Fig4}). Such highly non physical behavior gives rise to an increased dissipation and decoherence dynamics, which in turn results in the significant overestimation of the diffusion process reported in the lower panel of Fig.~\ref{Fig5}.

\begin{figure}
\centering
\includegraphics*[width=7.2cm]{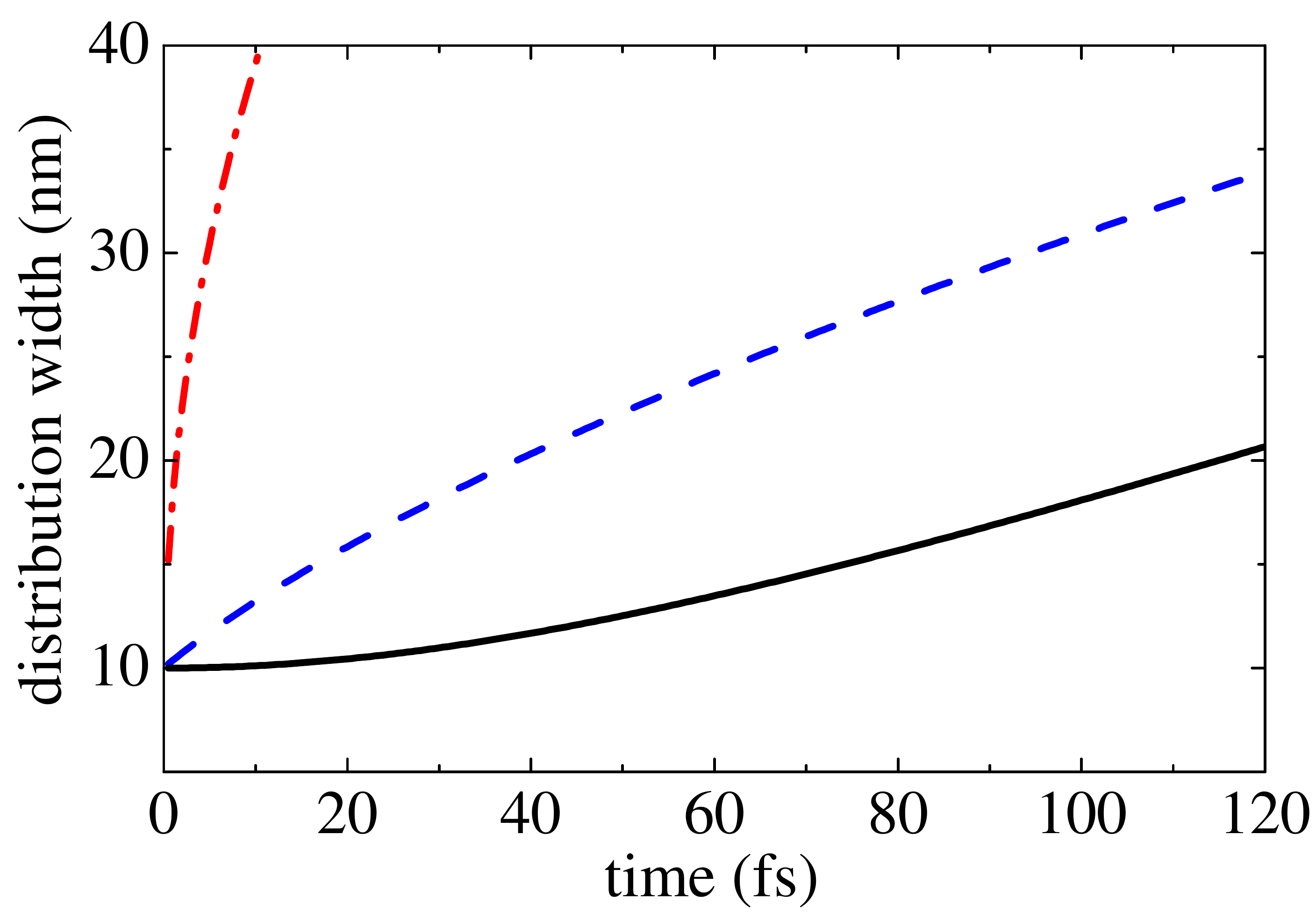}
\caption[]{(Color online)
Effective spatial-distribution width $\lambda$ in Eq.~(\ref{lambda}) as a function of time.
Here, the local-scattering result (see Eq.~(\ref{nzt})) (solid curve) is compared
to the corresponding results obtained adopting as scattering models the Lindblad superoperator in Eq.~(\ref{Lindbladalpha}) (dashed curve) as well as the relaxation-time model in Eq.~(\ref{RTA}) (dash-dotted curve) (see text).
}
\label{Fig6}       
\end{figure}

To quantify the amount of extra diffusion reported in Fig.~\ref{Fig5}, let us introduce the effective carrier distribution width
\begin{equation}\label{lambda}
\lambda = 
\sqrt{
\frac{
\int z^2 n(z)\, dz
}{
\int n(z)\,dz
}
}\ .
\end{equation}
Figure \ref{Fig6} shows the time evolution of the above effective distribution width $\lambda$.
Here, the local-scattering result $\lambda = \Delta_z(t)$ (solid curve) is compared
to the corresponding results obtained adopting as scattering models the Lindblad superoperator (\ref{Lindbladalpha}) (dashed curve) as well as the relaxation-time model (\ref{RTA}) (dash-dotted curve).
As expected, the relaxation-time model gives rise to a strong overestimation of the diffusion process (see dash-dotted curve) compared to the Lindblad-superoperator treatment (dashed curve).

As anticipated, the relaxation-time model in (\ref{RTA}) does not exhibit the well established in- minus out-scattering structure of the Boltzmann collision term as well as of the Lindblad superoperator in (\ref{Lindbladalpha}); it follows that within such simplified model the decay of the inter-state phase coherence (also referred to as inter-state polarization) is not dictated by a balance between in- and out-contributions, but is determined by out-scattering contributions only, leading to an overestimation of electronic decoherence.
In order to elucidate this crucial point, let us start by analyzing the explicit form of Eq.~(\ref{SBE}) for the case of the relaxation-time model in (\ref{RTA}). By denoting with
\begin{equation}\label{rhoi}
\rho_{\alpha_1\alpha_2}^{\rm i}(t) = \rho^{ }_{\alpha_1\alpha_2}(t) e^{-\frac{(\epsilon_{\alpha_1}-\epsilon_{\alpha_2}) t}{i\hbar}}
\end{equation}
the single-particle density matrix written in the interaction picture, the time evolution of its non-diagonal ($\alpha_1 \ne \alpha_2$) elements is given by
\begin{equation}\label{SBERTA}
\frac{d\rho_{\alpha_1\alpha_2}^{\rm i}}{dt}
=
-\,\frac{\Gamma_{\alpha_1} + \Gamma_{\alpha_2}}{2}\,\rho_{\alpha_1\alpha_2}^{\rm i}\ , 
\end{equation}
which shows that, in addition to the free rotation in (\ref{rhoi}), the inter-state polarization decays according to the decoherence rate $(\Gamma_{\alpha_1} + \Gamma_{\alpha_2})/2$.
In contrast, by inserting into Eq.~(\ref{SBE}) the explicit form of the Lindblad superoperator (\ref{Lindbladalpha}), it is easy to get
\begin{widetext}
\begin{equation}\label{SBELindblad}
\frac{d\rho_{\alpha_1\alpha_2}^{\rm i}}{dt}
=
\left(
\mathcal{L}_{\alpha_1\alpha_2,\alpha_1\alpha_2}
+
\mathcal{L}_{\alpha_2\alpha_1,\alpha_2\alpha_1}
\right)
\rho_{\alpha_1\alpha_2}^{\rm i}
+ 
\sum_{\alpha_1'\alpha_2' \ne \alpha_1\alpha_2}
\left(
e^\frac{(\epsilon_{\alpha_1'} - \epsilon_{\alpha_2'} - \epsilon_{\alpha_1} + \epsilon_{\alpha_2}) t}{i \hbar}
\mathcal{L}_{\alpha_1\alpha_2,\alpha_1'\alpha_2'}
\rho_{\alpha_1'\alpha_2'}^{\rm i}\, + \textrm{H.c.}\right)
\end{equation}
\end{widetext}
with
\begin{equation}\label{calL}
\mathcal{L}_{\alpha_1\alpha_2,\alpha_1'\alpha_2'} 
\!=\! 
\frac{1}{2} \sum_s
\left(
\mathcal{P}^s_{\alpha_1\alpha_2,\alpha_1'\alpha_2'} 
\!-\!
\delta_{\alpha_2\alpha_2'}
\sum_{\alpha'}
\mathcal{P}^{s *}_{\alpha'\alpha',\alpha_1\alpha_1'}
\right)\ .
\end{equation} 
In the presence of strongly nonelastic interaction processes, the overall impact of the second term in (\ref{SBELindblad}) is strongly reduced thanks to the fast temporal oscillations of the various free-rotation phase factors; moreover, taking into account that in such nonelastic-interaction limit $\mathcal{P}^s_{\alpha\alpha',\alpha\alpha'} \to 0$, one gets 
\begin{equation}\label{calLbis}
\mathcal{L}_{\alpha\alpha',\alpha\alpha'} \to -\Gamma_{\alpha}/2\ ,
\end{equation} 
which implies that in this limit the Lindblad-model equation in (\ref{SBELindblad}) reduces to the relaxation-time one in (\ref{SBERTA}). In contrast, in the presence of quasielastic processes one deals with a significant cancelation between in- and out-scattering contributions, not accounted for by the relaxation-time equation (\ref{SBERTA}).
It is worth stressing that such intrinsic limitation of relaxation-time models has been already recognized in the analysis of ultrafast phenomena in photoexcited semiconductors \cite{Rossi02b}, showing that the latter becomes particularly severe for the case of quasielastic processes.\cite{Rossi94c}

\begin{figure}
\centering
\includegraphics*[width=7.2cm]{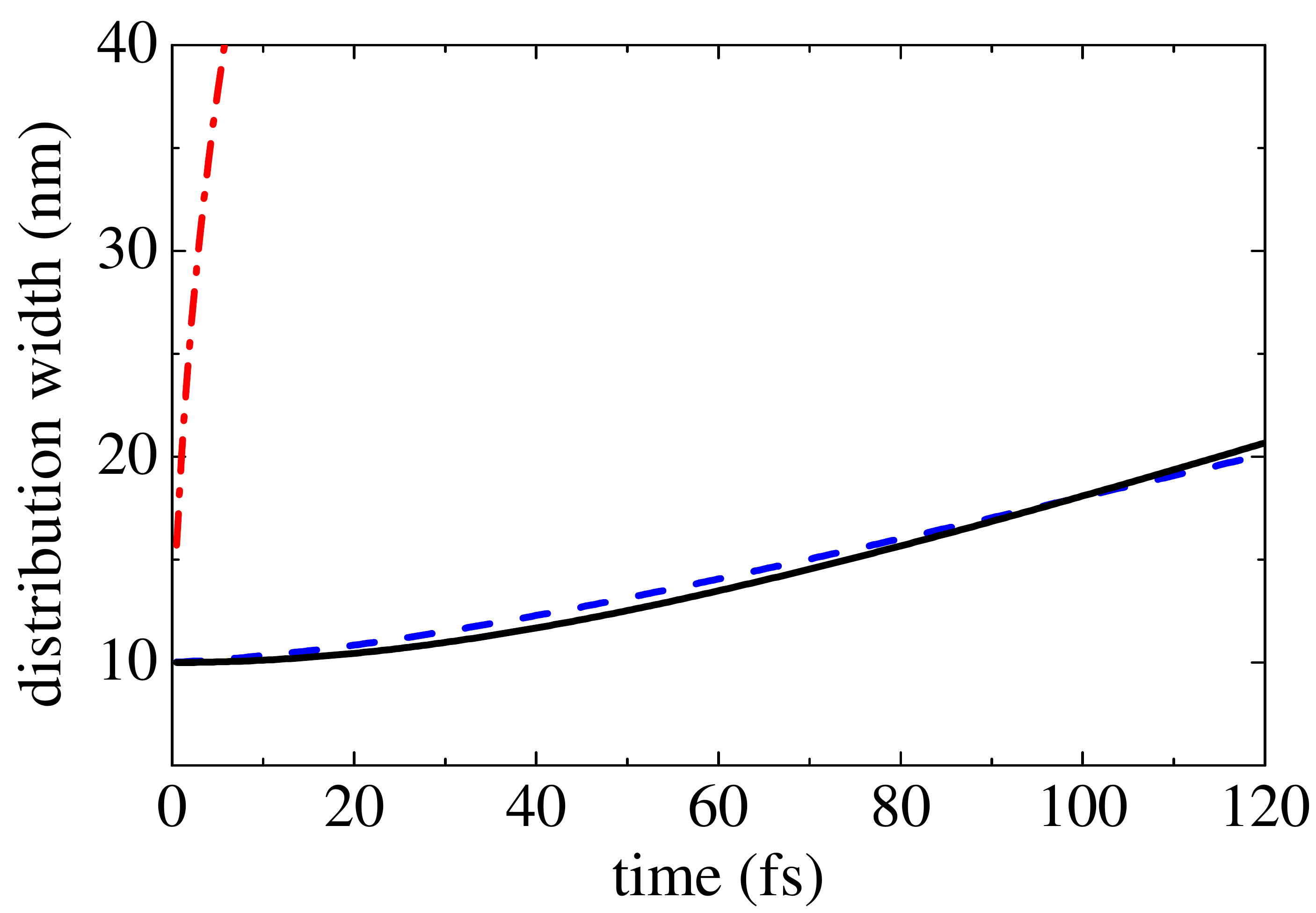}
\caption[]{(Color online)
Same as in Fig.~\ref{Fig6} but for a reduced value of the LO-phonon energy ($\epsilon_{\rm LO} = 20$\,meV) (see text).
}
\label{Fig7}       
\end{figure}

To confirm this physical interpretation, we have repeated the simulated experiments presented so far artificially reducing the GaN LO-phonon energy by a factor $4$ (from $80$ to $20$\,meV), such to mimic the quasielastic-process limit.
The time evolution of the effective distribution width $\lambda$ corresponding to these new simulations is reported in Fig.~\ref{Fig7}.
As expected, compared to the results reported in Fig.~\ref{Fig6}, the decoherence overestimation produced by the relaxation-time model (dash-dotted curve) is still increased, while the diffusion speed up induced by the Lindblad superoperator (dashed curve) is strongly reduced. Indeed, in spite of the fact that the LO-phonon energy is still significantly different from zero, the effect of phonon scattering is already negligible.
This is a clear indication that in the presence of genuine quasi elastic processes like, e.g., carrier-acoustic phonons or carrier-carrier scattering (i) the relaxation-time model is definitely inadequate, and (ii) quantum diffusion due to scattering nonlocality is expected to play a minor role.

\subsection{From homogeneous systems to nanostructures}\label{ss-fbstn}

\begin{figure}
\centering
\includegraphics*[width=7.2cm]{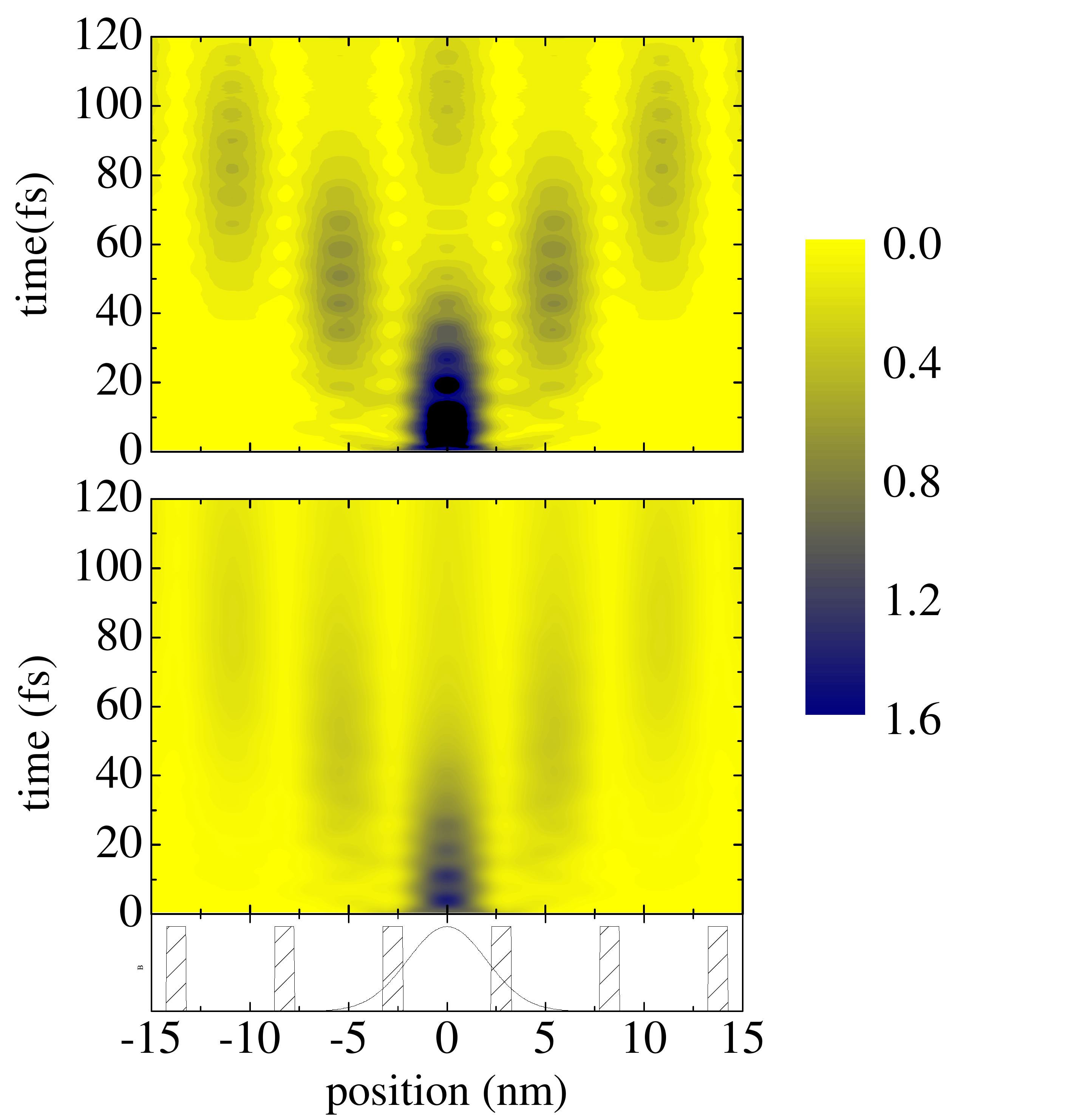}
\caption[]{(Color online)
Room-temperature quantum-diffusion dynamics in a GaN-based superlattice (lower panel) (band offset of $0.3$\,eV and well and barrier widths of $4.5$ and $1$\,nm) obtained in the absence of carrier-phonon coupling (upper panel) and via the Lindblad scattering superoperator in Eq.~(\ref{Lindbladalpha}) (central panel): sub-picosecond time evolution of the spatial carrier density corresponding to the initial mixed state in Eq.~(\ref{rhocirc}) with $\overline{\Delta}_z = 2$\,nm (see text).
}
\label{Fig8}
\end{figure}

As a final set of simulated experiments aimed at showing the power and flexibility of the proposed density-matrix approach, we have extended the homogeneous-system analysis presented so far to the case of a periodic nanostructure.
Figure \ref{Fig8} displays the sub-picosecond time evolution of the spatial carrier density in a GaN-based superlattice (see lower panel) corresponding to the initial mixed state in (\ref{rhocirc}) with $\overline{\Delta}_z = 2$\,nm, obtained in the scattering-free case (upper panel) and employing the Lindblad scattering superoperator in (\ref{Lindbladalpha}) (central panel).
Compared to the corresponding homogeneous-system results of Fig.~\ref{Fig5}, here the superlattice structure (see lower panel) gives rise to a non-trivial interplay between the spatial quantum confinement dictated by the nanostructure potential profile and the scattering-induced diffusion, resulting in a superlattice-induced modulation of the density profile.

\begin{figure}
\centering
\includegraphics*[width=7.2cm]{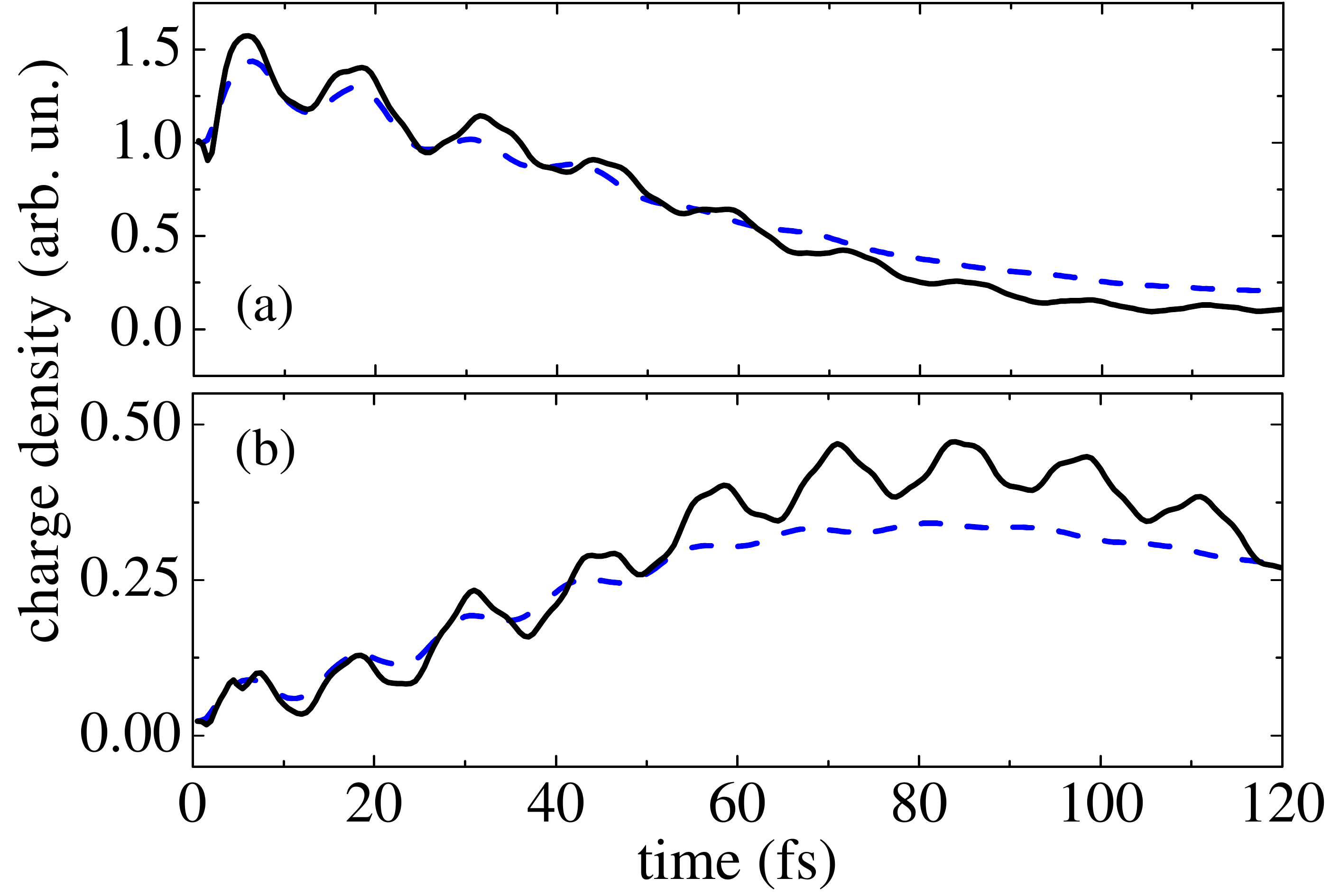}
\caption[]{(Color online)
Time evolution of the carrier population in the central well of the superlattice (panel a) as well as in the two adjacent wells (panel b) corresponding to the scattering-free simulation (solid curves (upper panel in Fig.~\ref{Fig8})) and to the Lindblad-scattering simulation (dashed curves (central panel in Fig.~\ref{Fig8})) (see text).
}
\label{Fig9}
\end{figure}

In the absence of carrier-phonon scattering (upper panel) one deals with coherent charge oscillations originating from the diffusion dynamics of the initial packet through the superlattice structure.
In particular, it is easy to recognize the typical signature of inter-well coherent tunneling, a peculiar phenomenon in coupled quantum-well structures.\cite{Iotti01b}
To better elucidate this crucial feature, in Fig.~\ref{Fig9} we have reported the time evolution of the carrier population in the central well of the superlattice (panel a) as well as in the two adjacent wells (panel b). 
As we can see, in the scattering-free case (solid curves corresponding to the upper-panel result of Fig.~\ref{Fig8}) one deals with a significant charge transfer
from the central well toward the adjacent ones and vice versa, the so-called coherent-tunneling dynamics. However, compared to simple two-well systems, here the situation is by far more complicated: once a fraction of the central-well charge has reached the adjacent wells, part of it will be transferred back to the central well, but also to the external nearest-neighbor ones; this process will progressively extend to an increasing number of wells, giving rise to the quantum-mechanical diffusion process displayed in the upper panel of Fig.~\ref{Fig8}.

In the presence of carrier-LO phonon scattering (see central panel in Fig.~\ref{Fig8} and dashed-curves in Fig.~\ref{Fig9}), the fully coherent dynamics just described is strongly suppressed; indeed, the significant temporal oscillations in Fig.~\ref{Fig9} are strongly reduced, giving rise at long times to a classical-like diffusion scenario typical of a so-called incoherent-tunneling dynamics.\cite{Iotti01b}

\begin{figure}
\centering
\includegraphics*[width=7.2cm]{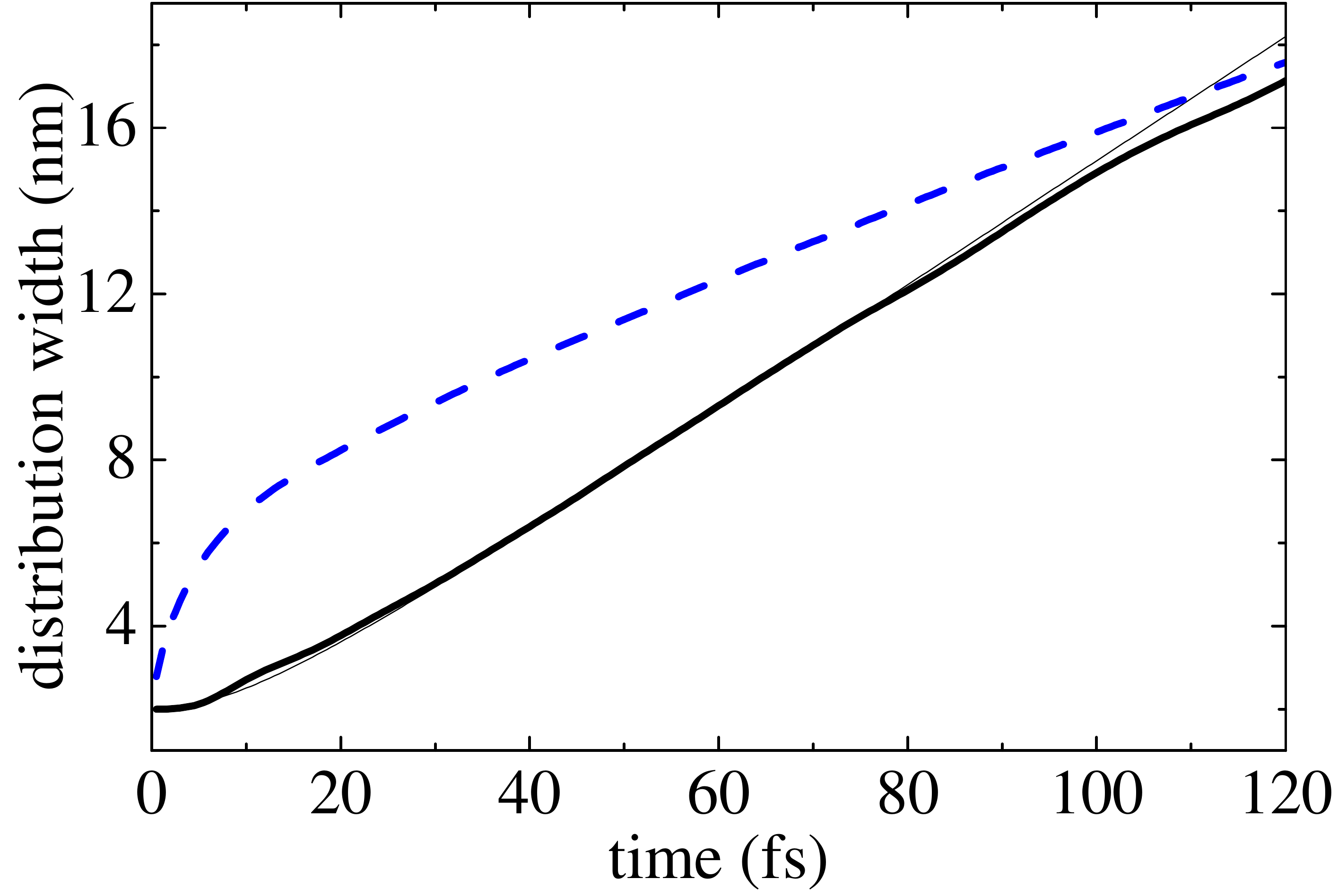}
\caption[]{(Color online)
Effective spatial-distribution width $\lambda$ in Eq.~(\ref{lambda}) as a function of time.
Here, the local-scattering homogeneous-system result in Eq.~(\ref{nzt}) (thin solid curve) is compared
to the scattering-free superlattice result (solid curve corresponding to the upper-panel result of Fig.~\ref{Fig8}) as well as 
to the Lindblad-scattering superlattice result (dashed curve corresponding to the central-panel result of Fig.~\ref{Fig8}) (see text).
}
\label{Fig10}       
\end{figure}

Finally, it is important to point out that in the presence of energy dissipation the interplay between single-particle phase coherence (dictated by the superlattice potential profile) and phonon-induced decoherence (dictated by the Lindblad scattering superoperator) is highly non trivial.
This is clearly shown in Fig.~\ref{Fig10}, where we report the effective spatial-distribution width $\lambda$ in (\ref{lambda}) corresponding to the two simulated experiments of Fig.~\ref{Fig8} as well as to the scattering-free homogeneous-system result of Fig.~\ref{Fig5}.

As we can see, at short times (less than $100$\,fs) the scattering-free diffusion dynamics within the superlattice structure (solid curve) does not differ significantly from the corresponding homogeneous-system result (thin solid curve).
In contrast, the presence of carrier-LO phonon scattering (dashed curve) gives rise to a significant diffusion speed up (compared to the scattering-free result (solid curve)); at longer times the non-local action of the scattering superoperator vanishes, and at the end of the simulation the spatial broadening induced by the Lindblad superoperator comes out to be similar to the scattering-free one.
Such non-trivial behavior can be explained as follows: at short times the strong spatial localization of the initial distribution induces a significant diffusion speed up due to carrier-phonon nonlocality effects; at longer times such scattering-induced nonlocality is strongly reduced, and, at the same time, energy dissipation tends to destroy inter-state phase coherence, thus limiting the diffusion process compared to the scattering-free case.

Generally speaking, we finally stress that the ability of investigating such space dependent phenomena originating from the complex interplay between single-particle quantum coherence and phonon-induced energy dissipation versus decoherence 
---definitely not possible via Boltzmann-like Monte Carlo simulation schemes--- constitutes a distinguished feature of the proposed quantum mechanical treatment.

\section{Summary and conclusions}\label{s-SC}

In this paper we have provided a rigorous treatment of scattering-induced spatial nonlocality in bulk as well as in nanostructured materials. 

On the one hand, starting from the conventional density-matrix formalism and employing as ideal instrument for the study of the semiclassical limit the well-known Wigner-function picture, we have performed a fully quantum-mechanical derivation of the space-dependent Boltzmann equation. 

On the other hand, we have analyzed the validity limits of such semiclassical approximation scheme, pointing out, in particular, regimes where scattering-nonlocality effects may play a relevant role; to this end we have supplemented our analytical investigation with a relevant set of simulated experiments, discussing and further expanding preliminary studies of scattering-induced quantum diffusion in GaN-based nanomaterials recently presented in Ref.~\onlinecite{Rosati14a}.

Our numerical investigation of ultrafast space-dependent phenomena in homogeneous GaN systems allows one to draw the following conclusions. 

In the presence of carrier localization on the nanometric space scale (see Fig.~\ref{Fig1}) within the proposed Lindblad treatment one deals with significant phonon-induced nonlocality effects; our analysis has shown that such non-local character is the result of a different spatial localization of in- and out-scattering contributions (see Figs.~\ref{Fig2} and \ref{Fig3});
these nonlocality effects will progressively vanish as the carrier delocalization increases, thus recovering, as expected, the local character of the Boltzmann collision term.
 
A detailed comparison of the proposed Lindblad scattering model (see Fig.~\ref{Fig1}) with the conventional relaxation-time approximation (see Fig.~\ref{Fig4}), has shown that the latter (i) leads to a significant overestimation of phonon-induced decoherence as well as scattering nonlocality, and (ii) is intrinsically unable to reproduce the local character of the Boltzmann collision term.

Thanks to our time-dependent simulations, we have shown that in homogeneous GaN systems one deals with a relevant competition between free-particle diffusion and phonon-induced non-local effects, giving rise to a global diffusion speed up (see Fig.~\ref{Fig5}); once again, a comparison between the proposed Lindblad treatment and the relaxation-time model has clearly shown that the latter leads to a significant overestimation of such diffusion speed up (see Fig.~\ref{Fig6}), and that this limitation is particularly severe for the case of quasielastic dissipation processes (see Fig.~\ref{Fig7}).

Moving from homogeneous systems to periodically modulated nanostructures, the interpretation of the diffusion process in the presence of phonon-induced dissipation is by far more complicated. Indeed, compared to the homogeneous-system results (see Fig.~\ref{Fig5}), the presence of the superlattice structure (see Figs.~\ref{Fig8} and \ref{Fig9}) gives rise to a non-trivial interplay between the spatial quantum confinement dictated by the nanostructure potential profile and the scattering-induced diffusion, resulting in a superlattice-induced modulation of the density profile.

Let us finally stress that in the presence of particularly strong interaction mechanisms as well as of extremely short electromagnetic excitations, the application of the Markov limit becomes questionable;\cite{Rossi02b,Axt04a} however, for a wide range of nanodevices and operation conditions the proposed Markov treatment is expected to well reproduce the sub-picosecond dynamics induced by a large variety of single-particle scattering mechanisms.

\appendix

\section{The semiclassical limit: Quantum-mechanical derivation of the Boltzmann collision term}\label{App-SL}

In order to derive the conventional Boltzmann collision term, the first step is to rewrite the Wigner scattering superoperator in (\ref{GammaWFbis}) within the momentum representation.
More specifically, denoting with
\begin{equation}\label{Ap1p2}
A^s(\mathbf{p}_1,\mathbf{p}_2) = \langle \mathbf{p}_1 \vert \hat A^s \vert \mathbf{p}_2 \rangle
\end{equation}
the (continuous) matrix elements of the Lindblad operators in (\ref{Lindblad}) and taking into account that
\begin{equation}\label{Wp}
\langle \mathbf{p}_1 \vert \hat{W}(\mathbf{r},\mathbf{p}) \vert \mathbf{p}_2 \rangle 
=
e^{\frac{(\mathbf{p}_1-\mathbf{p}_2) \cdot \mathbf{r}}{i\hbar}}
\delta\left(\frac{\mathbf{p}_1+\mathbf{p}_2}{2} - \mathbf{p}\right)\ ,
\end{equation}
the explicit form of the scattering superoperator in (\ref{GammaWFbis}) comes out to be
\begin{widetext}
\begin{eqnarray}\label{GammaWFter}
\Gamma(\mathbf{r},\mathbf{p};\mathbf{r}',\mathbf{p}')
&=& 
\left(\frac{2}{\pi\hbar}\right)^3
\sum_s \int d\mathbf{p}_1 d\mathbf{p}_2
e^{\frac{2(\mathbf{p}_1-\mathbf{p}_2+\mathbf{p}'-\mathbf{p}) \cdot \mathbf{r}}{i\hbar}}
A^s(2\mathbf{p}-\mathbf{p}_1, 2\mathbf{p}'-\mathbf{p}_2)
A^{s *}(\mathbf{p}_1,\mathbf{p}_2)
e^{-\frac{2(\mathbf{p}_2-\mathbf{p}') \cdot (\mathbf{r}'-\mathbf{r})}{i\hbar}}
\nonumber \\
&-& 
\left(\frac{2}{\pi\hbar}\right)^3
\sum_s \Re\left\{\int d\mathbf{p}_1 d\mathbf{p}_2
e^{\frac{2(\mathbf{p}'-\mathbf{p}) \cdot \mathbf{r}}{i\hbar}}
A^{s *}(\mathbf{p}_2,2\mathbf{p}-\mathbf{p}_1)
A^s(\mathbf{p}_2, 2\mathbf{p}'-\mathbf{p}_1)
e^{-\frac{2(\mathbf{p}_1-\mathbf{p}') \cdot (\mathbf{r}'-\mathbf{r})}{i\hbar}}
\right\}
\ .
\end{eqnarray} 
\end{widetext}
By inserting the above result into Eq.~(\ref{WE2Gamma}), one gets:
\begin{widetext}
\begin{eqnarray}\label{WE2Gammabis}
\left.\frac{\partial f^{\rm W}(\mathbf{r}\!,\!\mathbf{p})}{\partial t}\right|_{\rm scat}
\!&=&\! 
\left(\frac{2}{\pi\hbar}\right)^3
\sum_s 
\int d\mathbf{r}' d\mathbf{p}' d\mathbf{p}_1 d\mathbf{p}_2
e^{\frac{2(\mathbf{p}_1-\mathbf{p}_2+\mathbf{p}'-\mathbf{p}) \cdot \mathbf{r}}{i\hbar}}
A^s(2\mathbf{p}\!-\!\mathbf{p}_1, 2\mathbf{p}'\!-\!\mathbf{p}_2)
A^{s *}(\mathbf{p}_1,\mathbf{p}_2)
e^{-\frac{2(\mathbf{p}_2-\mathbf{p}') \cdot (\mathbf{r}'-\mathbf{r})}{i\hbar}}
f^{\rm W}(\mathbf{r}'\!,\!\mathbf{p}')
\nonumber \\
\!&-&\! 
\left(\frac{2}{\pi\hbar}\right)^3
\sum_s \Re\left\{
\int d\mathbf{r}' d\mathbf{p}' d\mathbf{p}_1 d\mathbf{p}_2
e^{\frac{2(\mathbf{p}'-\mathbf{p}) \cdot \mathbf{r}}{i\hbar}}
A^{s *}(\mathbf{p}_2,2\mathbf{p}\!-\!\mathbf{p}_1)
A^s(\mathbf{p}_2, 2\mathbf{p}'\!-\!\mathbf{p}_1)
e^{-\frac{2(\mathbf{p}_1-\mathbf{p}') \cdot (\mathbf{r}'-\mathbf{r})}{i\hbar}}
f^{\rm W}(\mathbf{r}'\!,\!\mathbf{p}')
\right\}
\ .
\nonumber \\
\end{eqnarray} 
\end{widetext}

Let us now analyze the semiclassical limit of the above quantum-mechanical scattering superoperator.
From a physical point of view, in the limit $\hbar \to 0$ the various phase factors entering Eq.~(\ref{WE2Gammabis}) will display infinitely fast oscillations, which allows one to evaluate some of the above coordinate and momentum integrals via a sort of adiabatic-decoupling procedure.
As far as the coordinate $\mathbf{r}'$ is concerned, for any regular function $F(\mathbf{r})$ we have:
\begin{equation}\label{lim1}
\lim_{\hbar \to 0} \int d\mathbf{r}' e^{\frac{\mathbf{p}'' \cdot (\mathbf{r}'-\mathbf{r})}{i\hbar}} F(\mathbf{r}') = (2\pi\hbar)^3 \delta(\mathbf{p}'') F(\mathbf{r})\ .
\end{equation}
By employing this general property, in the semiclassical limit ($\hbar \to 0$) the scattering superoperator in (\ref{WE2Gammabis}) simplifies to:
\begin{widetext}
\begin{eqnarray}\label{WE2Gammater}
\left.\frac{\partial f^{\rm W}(\mathbf{r},\mathbf{p})}{\partial t}\right|_{\rm scat}
&=& 
8
\sum_s 
\int d\mathbf{p}' d\mathbf{p}_1
e^{\frac{2(\mathbf{p}_1-\mathbf{p}) \cdot \mathbf{r}}{i\hbar}}
A^s(2\mathbf{p}-\mathbf{p}_1, \mathbf{p}')
A^{s *}(\mathbf{p}_1,\mathbf{p}')
f^{\rm W}(\mathbf{r},\mathbf{p}')
\nonumber \\
&-& 
8
\sum_s \Re\left\{
\int d\mathbf{p}' d\mathbf{p}_2
e^{\frac{2(\mathbf{p}'-\mathbf{p}) \cdot \mathbf{r}}{i\hbar}}
A^{s *}(\mathbf{p}_2,2\mathbf{p}-\mathbf{p}')
A^s(\mathbf{p}_2, \mathbf{p}')
f^{\rm W}(\mathbf{r},\mathbf{p}')
\right\}
\ .
\end{eqnarray} 
\end{widetext}
In addition to the spatial adiabatic decoupling in (\ref{lim1}), in the semiclassical limit it is also possible to show that for any regular function $G(\mathbf{r},\mathbf{p})$:
\begin{equation}\label{lim2}
\lim_{\hbar \to 0} \int d\mathbf{p}'' e^{\frac{(\mathbf{p}''-\mathbf{p}) \cdot \mathbf{r}}{i\hbar}} G(\mathbf{r},\mathbf{p}'') = \frac{(2\pi\hbar)^3}{\Omega} G(\mathbf{r},\mathbf{p})\ .
\end{equation}
Here $\Omega$ denotes a proper crystal normalization volume; indeed, in order to derive this result it is crucial to perform a sort of spatial coarse graining, i.e., a spatial average of the function $G$ over a volume $\Omega$ much larger than the typical carrier coherence length and much smaller than the macroscopic spatial variations of our material.
By employing the general property in (\ref{lim2}) the scattering superoperator in (\ref{WE2Gammater}) reduces to:
\begin{widetext}
\begin{equation}\label{WE2Gammaqua}
\left.\frac{\partial f^{\rm W}(\mathbf{r},\mathbf{p})}{\partial t}\right|_{\rm scat}
= 
\frac{(2\pi\hbar)^3}{\Omega}
\sum_s 
\int d\mathbf{p}' 
\left[
\left|A^s(\mathbf{p},\mathbf{p}')\right|^2
f^{\rm W}(\mathbf{r},\mathbf{p}')
- 
\left|A^s(\mathbf{p}',\mathbf{p})\right|^2
f^{\rm W}(\mathbf{r},\mathbf{p})
\right]
\ .
\end{equation} 
 \end{widetext}
This is exactly the Boltzmann collision term of the semiclassical theory we were looking for; indeed, the latter can be written in a more compact form according to Eq.~(\ref{bte4WF}), where
\begin{equation}\label{P}
P(\mathbf{p},\mathbf{p}') = \sum_s P^s(\mathbf{p},\mathbf{p}')
\end{equation}
and
\begin{equation}\label{PsWF}
P^s(\mathbf{p},\mathbf{p}') = 
\frac{(2\pi\hbar)^3}{\Omega}
\left|A^s(\mathbf{p},\mathbf{p}')\right|^2\ .
\end{equation}
This shows that the scattering rates of the Boltzmann transport theory can be easily expressed in terms of the matrix elements of the various Lindblad operators.

In order to establish a direct link with the conventional Fermi's-golden-rule prescription, let us finally move from the continuous momentum representation employed so far to its discrete version corresponding to the crystal normalization volume $\Omega$; more precisely, employing the usual continuous-versus-discrete prescription, the scattering rates in (\ref{PsWF}) can also be written as
\begin{equation}\label{PsWFbis}
P^s_{\mathbf{p},\mathbf{p}'} = 
\left|A^s_{\mathbf{p},\mathbf{p}'}\right|^2\ ,
\end{equation}
in total agreement with the diagonal-approximation result in (\ref{sl3}).

\section{Microscopic derivation of the scattering superoperator}\label{App-LSS}

Aim of this appendix is to briefly recall the basic steps and main results of the alternative adiabatic-decoupling approach proposed in Ref.~\onlinecite{Taj09b}.

Within the spirit of the usual perturbation theory, the global semiconductor Hamiltonian (electrons plus various quasi-particle excitations, e.g., phonons, plasmons, etc.) may be written as the sum of a so-called unperturbed contribution $\hat H_\circ$ which can be treated exactly, plus a perturbation term $\hat H'$ which is typically treated within some approximation scheme.
More specifically, by introducing a properly designed adiabatic-decoupling prescription (based on a time symmetrization between microscopic and macroscopic scales), it is possible to express the second-order (or scattering) contribution to the time evolution of the global density-matrix operator $\hat{\mathbf{\rho}}$
in terms of the Lindblad superoperator
\begin{equation}\label{Lindbladglobal}
\left.\frac{d \hat{\mathbf{\rho}}}{d t}\right|_{\rm scat}
= 
\hat{\mathbf{A}}^{ } \hat{\mathbf{\rho}} \hat{\mathbf{A}}^\dagger
- 
\frac{1}{2} \left\{\hat{\mathbf{A}}^\dagger \hat{\mathbf{A}}^{ }, \hat{\mathbf{\rho}}\right\}\ ,
\end{equation}
where
\begin{equation}\label{hatbfA}
\hat{\mathbf{A}} 
= 
\lim_{\overline{\epsilon} \to 0}
\left({2 \overline{\epsilon}^2 \over \pi\hbar^6}\right)^{1 \over
4} \int_{-\infty}^\infty dt' \hat{H}^{\prime {\rm i}}(t') e^{-\left({\overline{\epsilon} t' \over \hbar}\right)^2} 
\end{equation}
and
\begin{equation}\label{Hprimei}
\hat{H}^{\prime {\rm i}}(t) = 
e^{{\hat H_\circ t \over i \hbar}}
\hat H'
e^{-{\hat H_\circ t \over i \hbar}}
\end{equation}
is the perturbation Hamiltonian $\hat H'$ written in the interaction picture.\footnote{The energy $\overline{\epsilon}$ can be regarded as a sort of level broadening corresponding to a finite collision duration and/or to a finite single-particle life-time.\cite{b-Rossi11}}

We stress that, opposite to standard master-equation formulations,\cite{b-Davies76,Spohn80a} in this new adiabatic-decoupling strategy positivity is intrinsic, and does not depend on the chosen subsystem of interest; moreover, the above Markov prescription is valid regardless of the specific form of the interaction Hamiltonian $\hat H'$.

Starting from such global description, it is possible to derive an effective scattering superoperator within the usual single-particle picture (see Sec.~\ref{s-DMF}).
More specifically, by denoting with
\begin{equation}\label{rhoalpha}
\rho_{\alpha_1\alpha_2} 
= 
\langle \alpha_1 \vert \hat\rho \vert \alpha_2 \rangle 
=
{\rm tr}\left\{\hat c^\dagger_{\alpha_2} \hat c^{ }_{\alpha_1} \hat{\mathbf{\rho}}\right\}
\end{equation}
the single-particle density matrix
($\hat{c}^\dagger_\alpha$ and $\hat{c}^{ }_\alpha$ denoting the usual creation and destruction operators over the single-particle states $\vert \alpha \rangle$)
and employing the usual mean-field approximation,\cite{Rossi02b} for any single-particle interaction mechanism it is possible to derive the non-linear scattering superoperator in (\ref{NSS}), where the explicit form of the Lindblad operators $\hat A^s$ depends on the particular form of the interaction Hamiltonian $\hat H'$; moreover, by neglecting so-called Pauli factors, the latter reduces to the Lindblad scattering superoperator in (\ref{Lindblad}).

For the case of the carrier-phonon coupling considered in this paper, the noninteracting Hamiltonian is the sum of the electron and phonon contributions,
\begin{equation}\label{H0}
\hat H_\circ = 
\sum_\alpha \epsilon_\alpha \hat c^\dagger_\alpha \hat c^{ }_\alpha
+
\sum_\mathbf{q} \epsilon_\mathbf{q} \hat b^\dagger_\mathbf{q} \hat b^{ }_\mathbf{q} \ ,
\end{equation}
($\hat b^\dagger_\mathbf{q}$ and $\hat b^{ }_\mathbf{q}$ denoting creation and destruction of a phonon with wavevector $\mathbf{q}$ and energy $\epsilon_\mathbf{q}$),
while the interaction Hamiltonian is given by
\begin{equation}\label{Hprimecp}
\hat H' = \sum_{\alpha\alpha',\mathbf{q}}
\left(
g^{\mathbf{q} -}_{\alpha\alpha'}
\hat c^\dagger_{\alpha} \hat b^{ }_\mathbf{q} \hat c^{ }_{\alpha'}
+
g^{\mathbf{q} +}_{\alpha\alpha'} \hat c^\dagger_{\alpha'}\hat b^\dagger_\mathbf{q} \hat c^{ }_{\alpha} \right) \ ,
\end{equation}
where
$
g^{\mathbf{q} \pm}_{\alpha\alpha'} = g^{\mathbf{q} \mp *}_{\alpha'\alpha}
$
are carrier-phonon matrix elements for the single-particle transition $\alpha' \to \alpha$ induced by the phonon mode $\mathbf{q}$, whose explicit form depends
on the particular interaction mechanism under examination (for the carrier-LO phonon coupling considered in our simulated experiments the latter scale as $q^{-1}$).

In this case the generic electron dissipation channel corresponds to the emission ($+$) or absorption ($-$) of a phonon with wavevector $\mathbf{q}$ and energy $\epsilon_\mathbf{q}$, i.e., $s \equiv \mathbf{q} \pm$, and the Lindblad scattering superoperator in (\ref{Lindblad}) comes out to be
\begin{equation}\label{Lindbladcp}
\Gamma \,(\hat{\rho})
=
\sum_{\mathbf{q} \pm} \left(
\hat A^{\mathbf{q} \pm} \hat\rho \hat A^{\mathbf{q} \pm \dagger}
- 
\frac{1}{2} \left\{\hat A^{\mathbf{q} \pm \dagger} \hat A^{\mathbf{q} \pm}, \hat\rho\right\} 
\right) \ ,
\end{equation}
where the matrix elements of the carrier-phonon Lindblad operators $\hat A^{\mathbf{q} \pm}$ are given by
\begin{equation}\label{Aalphacp}
A^{\mathbf{q} \pm}_{\alpha\alpha'} =
\sqrt{
2 \pi \left(N_\mathbf{q}+{1 \over 2} \pm {1 \over 2}\right)
\over 
\hbar
}
g^{\mathbf{q} \pm}_{\alpha\alpha'} D^{\mathbf{q} \pm}_{\alpha\alpha'}
\end{equation}
with
\begin{equation}
D^{\mathbf{q} \pm}_{\alpha\alpha'} =
\lim_{\overline{\epsilon} \to 0}
{
e^{-\left({\epsilon_\alpha-\epsilon_{\alpha'} \pm \epsilon_\mathbf{q} \over 2
\overline{\epsilon}}\right)^2 }
\over \left(2\pi \overline{\epsilon}^2\right)^{{1 \over 4}}
} \ .
\end{equation}
By inserting the explicit form of the matrix elements in (\ref{Aalphacp}) into Eq.~(\ref{calP}), the explicit form of the generalized carrier-phonon scattering rates comes out to be
\begin{widetext}
\begin{equation}\label{calPcp}
\mathcal{P}_{\alpha_1\alpha_2,\alpha'_1\alpha'_2} 
=
\lim_{\overline{\epsilon} \to 0}
{2\pi \over \hbar}
\sum_{\mathbf{q} \pm}
\left(N_\mathbf{q}+{1 \over 2} \pm {1 \over 2}\right)
g^{\mathbf{q} \pm}_{\alpha_1\alpha_1'} 
g^{\mathbf{q} \pm *}_{\alpha_2\alpha_2'} 
{
e^{-\left({\epsilon_{\alpha_1}-\epsilon_{\alpha_1'} \pm \epsilon_\mathbf{q} \over 2
\overline{\epsilon}}\right)^2 }
e^{-\left({\epsilon_{\alpha_2}-\epsilon_{\alpha_2'} \pm \epsilon_\mathbf{q} \over 2
\overline{\epsilon}}\right)^2 }
\over 
\left(2\pi \overline{\epsilon}^2\right)^{{1 \over 2}}
} 
\ .
\end{equation}
\end{widetext}

\begin{acknowledgments}

We are extremely grateful to David Taj and Rita Claudia Iotti for stimulating and fruitful discussions.

We gratefully acknowledge funding by the Graphene@PoliTo laboratory of the Politecnico di Torino, operating within the European FET-ICT Graphene Flagship project (www.graphene-flagship.eu).

Computational resources were provided by HPC@PoliTo, a project of Academic Computing of the Politecnico di Torino (www.hpc.polito.it).

\end{acknowledgments}


%

\end{document}